\documentclass[preprint,superscriptaddress,amsmath,amssymb,aps,physrev,prb,floatfix, fleqn]{revtex4-2} \usepackage{bm} 
\usepackage{hyperref}
\usepackage[english]{babel}
\usepackage[utf8]{inputenc}   
\usepackage{xspace}
\usepackage{tensor} 
\usepackage{array,color}
\usepackage{relsize,exscale}  
\usepackage{textcase}	
\usepackage{soul,color}		
\usepackage{calc} 
\usepackage{makeidx,showidx}
\usepackage{graphicx}
\usepackage{wrapfig}
\usepackage{textcomp}
\usepackage{siunitx}
\usepackage[version=3]{mhchem}
\usepackage{enumitem}

\newcommand{\erww} [1] {\ensuremath{\langle {#1} \rangle}}

\newcommand{\ybco} {$\ce{YBa2Cu3O_{6+y}}$\@\xspace}

\newcommand{\ybcoE} {$\ce{YBa2Cu4O8}$\@\xspace}

\newcommand{\tc} {\ensuremath{T_{\mathrm c}}\@\xspace}
\newcommand{\cperp}{\ensuremath{c \bot B_0}\@\xspace}
\newcommand{\cpara}{\ensuremath{{c\parallel\xspace B_0}}\@\xspace}
\newcommand{\tcmax} {\ensuremath{T_{\mathrm{c, max}}}\@\xspace}

\newcommand{\beq} {\begin{equation}}
\newcommand{\eeq} {\end{equation}}

\preprint{Lee NMR}
\begin{document}

\title{\textbf{Hidden Universal Metal in Cuprate Superconductors}}
\author{Abigail Lee}
\author{Jürgen Haase}
\email{juergen.haase@uni-leipzig.de}
\affiliation{University of Leipzig, Felix Bloch Institute for Solid State Physics, Linn\'estr.\@ 5, 04103 Leipzig, Germany}

\date{\today}
\medskip

\begin{abstract}
\vspace{1cm}
{Nuclear relaxation is a very robust probe of electronic excitations in superconducting materials, above and below the critical temperature of superconductivity, $T_\mathrm{c}$. Here, a phenomenology of it in cuprate superconductors is established based on essentially all cuprate data available in the literature from the CuO$_2$ plane. A universal 'hidden metal' with $1/T_1 T = const$ reigns below the pseudogap line, and, similar to usual superconducting metals, all cuprates condense at $T_\mathrm{c}$ out of this metal and relaxation ceases rapidly. There is no Hebel-Slichter peak. Above the hidden metal, a renormalized two-component metal is found. Therefore, the hidden metal is identified as the pseudogap matter, which has other important properties. It predominantly lacks a uniform response, unlike the normal metal above $T^*$, and it does not significantly relax planar O. However, for planar Cu it exhibits a special relaxation anisotropy. With the field in the plane, it causes a relaxation rate that is very similar for all cuprates, $1/{^{63}T}_{1\perp}T\approx 25/$Ks, while with the field parallel to the $c$-axis, $1/{^{63}T}_{1\parallel}T$, the proportional rate (in the pseudogap range) changes as a function of doping and material. This relaxation anisotropy is strictly correlated with the maximum critical temperature, $T_\mathrm{c,max}$, of all cuprates. It is stipulated that two partially independent spin components are needed to understand this behavior. The new phenomenology will be discussed and should give a better foundation for the understanding of the cuprates.}
\end{abstract}


\maketitle

\newpage

\section{Introduction}
Nuclear relaxation (1/$T_1$) is a powerful probe of the imaginary part of the electronic spin susceptibility \cite{Slichter1990}. In metals, it is proportional to temperature, $1/T_1 \propto T$ \cite{Heitler1936}, and the Korringa relation \cite{Korringa1950} links the NMR spin shift, $K$ (Pauli susceptibility), to relaxation; in simple scenarios,  $1/T_1 T = (\gamma_n/\gamma_e)^2 4\pi k_\mathrm{B}/\hbar \cdot K^2$.
 The Hebel-Slichter peak \cite{Hebel1957} demonstrates that even minute changes in the metallic electronic density of states due to the opening of a superconducting gap can be detected, before relaxation ceases deeper in the condensed state. This behavior still serves as a solid reference even when investigating complex modern materials with NMR.

With the advent of the superconducting cuprates \cite{Bednorz1986}, NMR investigations were in demand. Of greatest interest were magnetic shift and relaxation from $^{63,65}$Cu and $^{17}$O in the ubiquitous CuO$_2$ plane. These nuclei also have electric quadrupole moments, which lead to splittings of the nuclear resonances, and allow relaxation through lattice vibrations. This relaxation was reported to reach magnetic relaxation levels only for planar O in \ybcoE near \tc \cite{Suter2000}, but might not have been explored in detail for other materials (the expected temperature dependence from lattice effects can be complicated). While this electric interaction complicates NMR experiments in terms of sensitivity and resolution, it can also give vital information about the sharing of holes between Cu and O in the plane \cite{Zheng1995h,Jurkutat2014}, which is related to the maximum critical temperature of the cuprates \cite{Rybicki2016,Kowalski2021}

Most early experimental information on shift and relaxation came from measurements of the \ybco class of materials  \cite{Walstedt1988,Takigawa1989,Pennington1989,Zimmermann1989,Mila1989b}. They revealed a temperature dependent shift for Cu when the magnetic field ($B_0$) is in the plane, but not when the magnetic field is pointed in the crystal $c$-direction (\cpara), e.g.\@ in \ybcoE. With an isotropic spin response, this can only happen if the shift from spin in the Cu $3d(x^2-y^2)$ orbital (coupling coefficient $A_\parallel$, large and negative in related ionic materials) is offset accidentally by a second term. In a single band picture, it was then assumed that $A_\parallel + 4 B \approx 0$ and $A_\perp + 4B >0$, with transferred Cu spin density and isotropic coupling $B$ \cite{Mila1989b}. This hyperfine scenario also has profound consequences for understanding Cu nuclear relaxation, as it requires antiferromagnetic spin fluctuations along \cpara, as longer wavelengths are increasingly suppressed, preventing an unbiased view of the cuprates.


\begin{figure}[h!]
\centering
\includegraphics[width=.7\textwidth ]{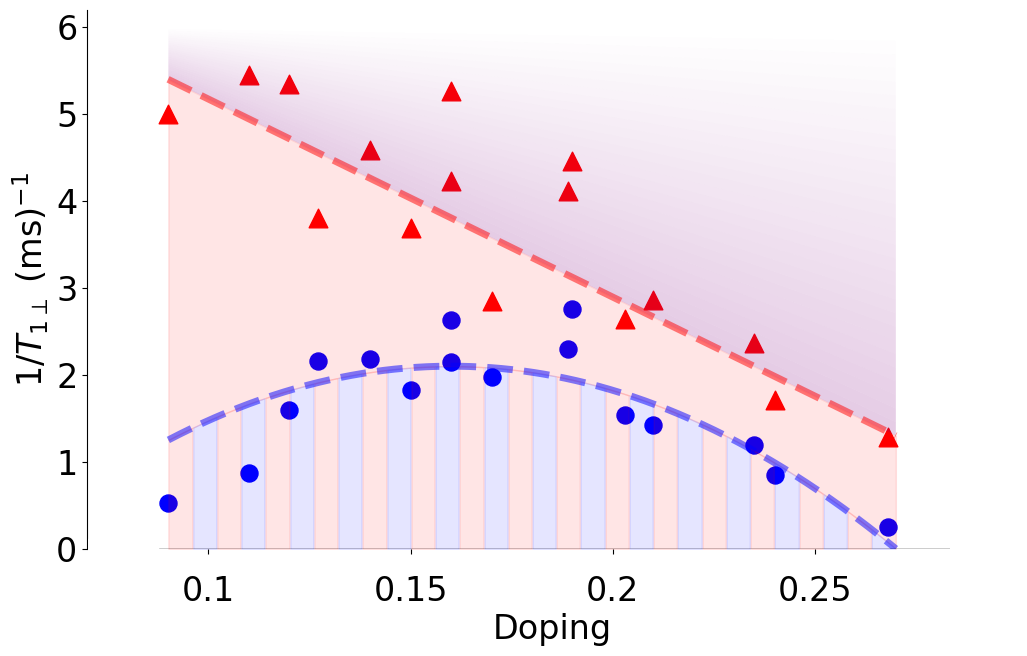}
\caption{Scaled nuclear relaxation rate $1/T_{1\perp}^*$ of planar Cu as a function of doping for many different cuprates. The relaxation is normalized to 1/$T_{1\perp}^*T=25$/Ks in the hidden metal regime (the actual variation is about $\sim15\%$ for most materials; see main text). Blue circles mark the relaxation rate with which a material enters this hidden metal regime above $T_\mathrm{c}$; red triangles mark the rate above which is departs from it, i.e.\@ where it begins to lag behind the universal metal relaxation. Both of these boundaries are defined by the temperature points at which the relaxation deviates by 6\% from the metallic rate. The red shaded area thus approximates the temperature-doping regime of this special metal. Within this regime, the relaxation anisotropy is temperature independent. Note that the condensate (red-blue striped region) always forms out of this metal.  Above the red triangles (approximated by the purple shaded region), we see a renormalized two-component metal. The dashed red line is a linear guide to the eye; the blue dashed parabola is an adaptation of the \tc-doping relation \cite{Presland1991}. Within the hidden regime, the relaxation rate can be directly translated into temperature, resulting in the similarity in appearance to the temperature-doping phase diagram. Note that the points are not normalized with respect to the family-dependent \tcmax, resulting in the scatter seen here.}\label{fig:fig1}
\end{figure}


As data from more cuprates appeared, it became increasingly obvious that for most materials, the Cu shift for \cpara is also strongly temperature dependent, overturning the old hyperfine scenario. Special tests of whether a single spin component view holds beyond \ybcoE, \ybco \cite{Takigawa1991,Bankay1994} showed that that is not the case \cite{Haase2009,Rybicki2015}. 

Very recently, using nearly all literature shift data, it was shown that there are indeed two electronic spin components A and B at work. One, A, stemming from the Cu $3d(x^2-y^2)$ spin, which sets the planar O as well as Cu axial shift, and another, B, Cu $4s$-like spin that affects only planar Cu \cite{Bandur2026}. This conclusion was confirmed  independent of the size of the corresponding hyperfine coefficients, the anisotropic $A_\alpha$ and isotropic $B$ \cite{Lee2026s}. 

Here we develop a phenomenology of nuclear relaxation based on the same symmetry considerations using nearly all available literature relaxation data \cite{Jurkutat2019,Nachtigal2020} (note that LSCO is excluded here due to its outlier behavior, including the presence of an additional axial shift component, \cite{Bandur2026} which complicates the analysis). Surprisingly, a simple, universal picture emerges in terms of the two spin components A and B.

At the center of the new phenomenology is the universal, 'hidden metal' that is present in all cuprates. 
It causes universal, metallic Cu relaxation of $1/{^{63}T}_{1\perp}T \approx 25$/Ks below about the pseudogap temperature down to the lowest temperatures (i.e. proportionality between rate and temperature is not broken while measured above \tc). At $T_\mathrm{c}$, a gap opens in the density of states and condensation into the superconducting state occurs out of this hidden metal. $1/{^{63}T}_{1\parallel}$ is proportional to $1/{^{63}T}_{1\perp}$ in the pseudogap range, but it is material and doping dependent. Surprisingly, $T_\mathrm{c}$ is a (single-valued) function of this relaxation anisotropy, which highlights the importance of the pseudogap matter. Planar O relaxation, quite to the contrary, is largely suppressed in the pseudogap as the O nucleus only coupes to the A-spin, unlike Cu. This behavior points to (short range) antiferromagnetic order of the pseudogap matter. This is also supported by the fact that the uniform response (spin shift) of the hidden metal is essentially zero, for A- as well as B-spins. Outside the hidden metal, simple A- and B-spin metals then set relaxation and shift in the conventional sense. 

This simple phenomenology should hold important clues for the microscopic picture of the cuprates.

\section{Nuclear Relaxation}

\begin{figure}
\centering
\includegraphics[width=.9\textwidth ]{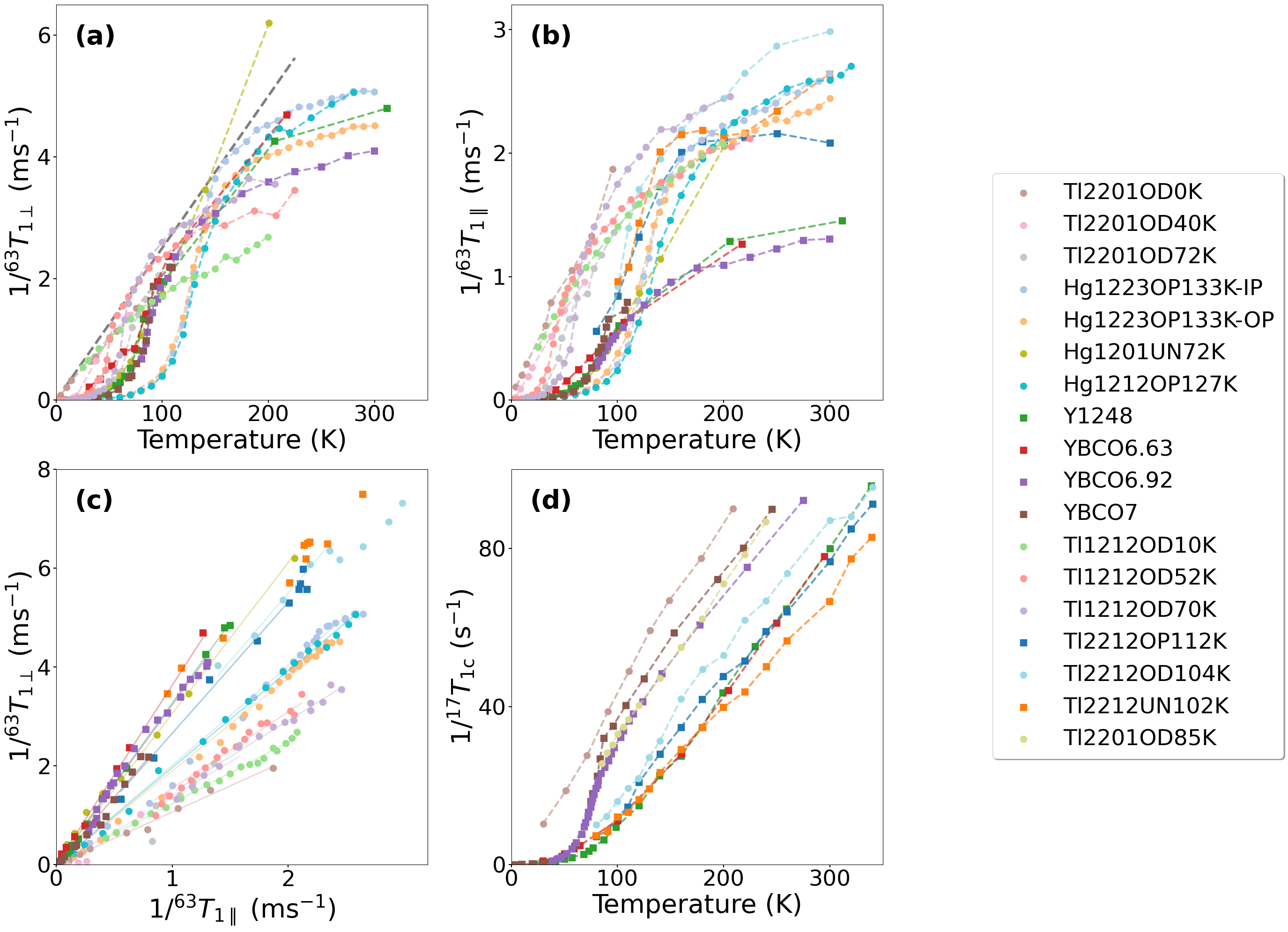}
\caption{Nuclear relaxation in a variety of hole-doped cuprates.  (a) 1/$^{63}T_{1\perp}$ vs $T$ (field applied perpendicular to the crystal c-axis). The grey dashed line marks a rate of 25/Ks.  (b) 1/$^{63}T_{1\parallel}$ vs $T$ (field applied along the crystal c-axis) shows a much larger variation between family and doping. (c) A plot of 1/${^{63}T}_{1\perp}$ vs 1/${^{63}T}_{1\parallel}$ with temperature as an implicit parameter. This relaxation anisotropy is material and doping dependent, but temperature independent within a certain window above \tc \cite{Jurkutat2019}. The variation in relaxation anisotropy is primarily set by the changes in 1/${^{63}T}_{1\parallel}$. (d) Planar oxygen relaxation follows what appears to be a simple metallic behavior, albeit with a gap in the low-energy DOS at lower doping levels \cite{Nachtigal2020}. Much of the data displayed for Cu and O alike have been presented cumulatively before \cite{Jurkutat2019, Nachtigal2020}. Original sources for this data as well as the additional data can be found in Table \ref{tab:table1} in the Appendix. }\label{fig:fig2}
\end{figure}

Planar Cu relaxation is either reported with the external magnetic field along the crystal $c$-axis (\cpara) or perpendicular to it (\cperp). The former measures $1/{^{63}T}_{1\parallel}$ (relaxation due to in-plane fluctuating fields, perpendicular to the $c$-axis, which are also measured in NQR); the latter measures $1/{^{63}T}_{1\perp}$ (due to both out-of-plane and in-plane fluctuating fields), independent of in-plane axis rotation due to symmetry. The anisotropy of nuclear relaxation, ${^{63}R}={^{63}T}_{1\parallel}/{^{63}T}_{1\perp}$, is expected to yield information about the anisotropy of the hyperfine coefficients, as the electronic spin is isotropic. A summary of typical experimental data is shown in Fig.~\ref{fig:fig2}(a, b), cf.~\cite{Jurkutat2019}. 


Before we explore the Cu relaxation in greater detail, we address the known phenomenology of planar {$^{17}$O} nuclear relaxation. It is known to be metallic with $1/{^{17}T}_{1\mathrm{c}}T \approx 0.40/$Ks beyond about optimal doping for all materials, cf.\@ Fig.~\ref{fig:fig2}(d) (for more data see \cite{Nachtigal2020}). The same metallic behavior is found even at lower doping levels, albeit at increasingly higher temperatures and with an offset, i.e.\@ it defines a temperature independent pseudogap. With the Korringa relation, a metallic shift of about 0.25\% follows, in excellent agreement with experiment \cite{Nachtigal2020}. The anisotropy (not shown) is largely material independent and expected by the hyperfine coefficients; therefore one can focus on the most abundant data for \cpara. Planar O thus points to a rather well-defined simple, universal metal in the cuprates that apparently condenses at \tc, except that below about optimal doping, it lacks low-energy states due to a temperature independent pseudogap, obscuring the condensation behavior. The planar O shifts show the same pseudogap behavior as relaxation, as states from the pseudogap are missing \cite{Nachtigal2020}.

Cu relaxation, panels (a-b), is very different. Salient features are the following. 
First, we notice that \tc is easily visible in the bare data for all materials in Fig.~\ref{fig:fig2}(a); it is also present for \cpara, although not as easily seen in Fig.~\ref{fig:fig2}(b). Apparently, the pseudogap behavior does not obscure the onset of condensation, as for O, and one can measure $T_1$ near $T_\mathrm{c}$ (blue points in Fig.~\ref{fig:fig1}). 

Second, $1/{^{63}T}_{1\perp}$ at \tc is found on very similar metallic lines, i.e.\@ $1/{^{63}T}_{1\perp} T_\mathrm{c}$ is rather similar for all cuprates (within 15\% for the majority). However, the relaxation measured along the crystal $c$-axis, $1/{^{63}T}_{1\parallel}$, varies significantly between materials. It has been noted before \cite{Jurkutat2019} that this relaxation anisotropy, cf.\@ Fig.~\ref{fig:fig2}(c) is temperature independent (within the shaded region of Fig. \ref{fig:fig1}), but does depend on doping and family. It has the tendency to become more isotropic at high doping levels, and can be as large as 3.6 for certain materials at lower doping levels.

\begin{figure}
\centering
\includegraphics[width=.7\textwidth ]{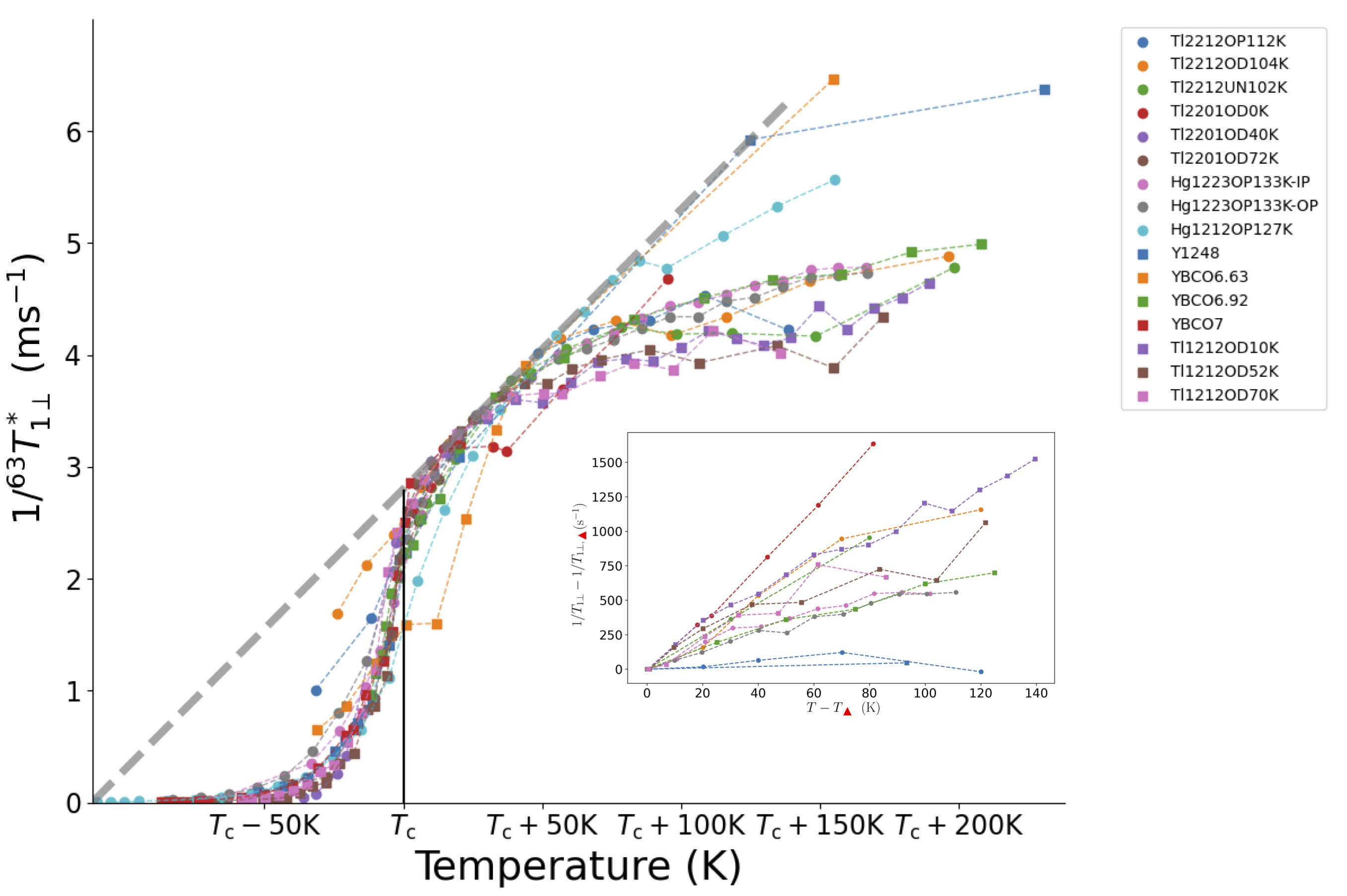}
\caption{Main panel: Scaled $1/{^{63}T}^*_{1\perp}$ plotted as a function of a shifted temperature axis such that the quoted \tc for each material is aligned to the black vertical line. $1/{^{63}T}_{1\perp}^*T$ ($1/{^{63}T}_{1\perp}T$ adjusted to fit $1/{^{63}T}_{1\perp}T=25$/Ks, dashed grey line in the figure) is multiplied by the shifted $T=T_\mathrm{c}$ (see main text for details). This serves to highlight the universality of the relaxation curves. Above \tc, materials follow this metallic line for a doping-dependent finite window in temperature before turning off. Note that given the proportionality, a similar curve can be obtained from  $1/{^{63}T}_{1\parallel}$ if additionally scaled by the observed relaxation anisotropies. Inset: $1/{^{63}T}_{1\perp}-1/{^{63}T}_{1\perp\textcolor{red}\blacktriangle}$ vs. $T-T_{\textcolor{red}\blacktriangle}$, where the red triangle denotes rate and temperature at the red triangles in Fig.~\ref{fig:fig1}, i.e.\@ at the onset of the hidden metal. In this regime, the relaxation anisotropy deviates from the temperature-independent values seen in the lower temperature regime. Dotted lines are guides to the eye.}\label{fig:fig3}
\end{figure}

Third, the range of temperatures over which the relaxation rate of a particular material stays on its metallic line is doping dependent. In order to investigate this in more detail, it is useful to normalize all metallic lines for \cperp, since slight differences in the slope obscure, in terms of the actual rate, how large the temperature range is over which a material follows the metallic slope. For that reason, we set the slopes to $1/{^{63}T}^*_{1\perp}T = 25$/Ks. It emphasizes the observation that the cuprates have very similar $1/{^{63}T}_{1\perp}T$, in particular despite the data coming from groups around the world, on all sorts of cuprates with different 'sample quality'. This rate is indicated by the dashed line in Fig.~\ref{fig:fig2}(a). This behavior is easily observed in Fig.~\ref{fig:fig3}, where we further normalize all measured $1/{^{63}T}^*_{1\perp}$ by the reported \tc, i.e. the temperature dependent rates, $1/{^{63}T}_{1\perp}T$, are shifted to fixed \tc and multiplied by \tc. Apart from the drop near $T_\mathrm{c}$, there is universal metallic behavior up  to $T_{\textcolor{red}\blacktriangle}$, cf.~Fig.~\ref{fig:fig1}.  Above $T_{\textcolor{red}\blacktriangle}$, the relaxation lags behind the Cu hidden metal rate. 

In the hidden metal regime, the relaxation anisotropy, ${^{63}R}$ varies with doping and family, but is independent of temperature \cite{Jurkutat2019}. Thus, while $1/{^{63}T}_{1\parallel}$ behaves differently from $1/{^{63}T}_{1\perp}$, we would find the same universal temperature dependence in this region of the phase diagram if we multiply $1/{^{63}T}_{1\parallel}$ by ${^{63}R}$. At higher temperatures, above the hidden metal line, the relaxation anisotropy decreases for all materials shown here. This high-temperature addition to the relaxation for \cperp is isolated in the inset of Fig. \ref{fig:fig3}. Note that the fixed anisotropy below the hidden metal line is not expected to strictly apply to the superconducting state, for which the relaxation behavior will be governed by a different functional dependence. 

\section{Discussion}
\begin{figure}[h!]
\centering
\includegraphics[width=.6\textwidth ]{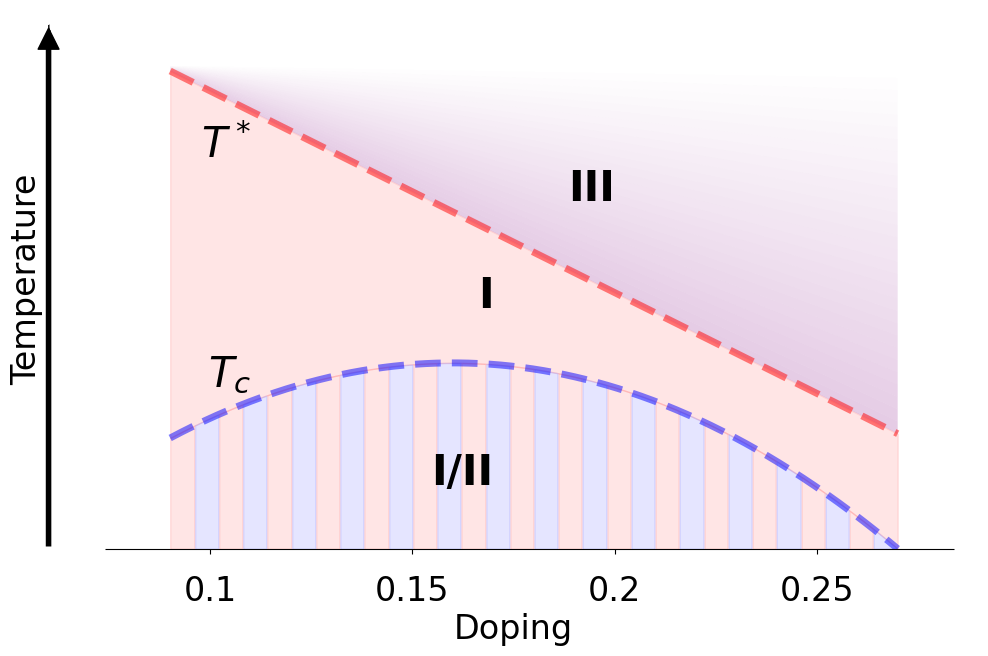}
\caption{Suggested generic cuprate phase diagram with phase boundaries defined by planar Cu nuclear relaxation. The hidden metal dominates region I (red). The striped region below the blue dashed line indicates the superconducting state (II) which forms out of this special metal. Region III is governed by a different, apparently renormalized two-component metal with temperature $T-T^*$ (see text for details). The hidden metal should persist down to $T=0$ were the superconducting state not to intervene, therefore we label the dome as region I/II.}\label{fig:fig4}
\end{figure}

Examining planar Cu  nuclear relaxation allows us to define three distinct regions in the cuprate phase diagram, as shown in Fig. \ref{fig:fig4}. In region I, the universal hidden metal relaxation persists, with small differences in this rate between families. This behavior (constant $1/T_{1}T$) is that which one expects for an ordinary metal. Within this region, the Cu relaxation anisotropy is temperature independent but family and doping dependent. We will discuss this anisotropy in greater detail later, as it encodes the superconducting critical temperature, \tc. 


As temperature is decreased further we enter region I/II. Near the superconducting transition at \tc (blue dashed line), we observe an abrupt and steep decrease in relaxation as the superconducting gap opens, and electronic states available to relax the nuclear spins disappear. In this sense the hidden metal behaves like a conventional superconducting metal, but is missing the Hebel-Slichter peak \cite{Hebel1957}. This is expected for predominantly d-wave pairing in conventional metals \cite{Schrieffer2007, Dai2024}. For the most overdoped sample, in the absence of \tc, the hidden metal remains from the red dashed line all the way down to the lowest temperature. 

With increasing temperature, we enter region III. The relaxation behavior changes abruptly from that of the hidden metal upon crossing the red dashed line. As was shown in the inset to Fig. \ref{fig:fig3}, the additional high temperature relaxation also looks essentially metallic, but it is a renormalized metal with temperature $T-T^*$. This will be discussed in further detail later.

The red dashed line in Fig. \ref{fig:fig4} either coincides with or precedes (with decreasing temperature or doping, so far as data are available) the $T^*$ line defined by the size of the gap in the density of states, as observed through the planar O shift \cite{Nachtigal2020} or entropy \cite{Loram1998}. This leads us to identify it as the matter that forms the pseudogap. It is not important whether a material is underdoped or overdoped, nor whether we speak of single-layer, bilayer, or multilayer materials; the superconducting state always forms out of this pseudogap matter when the superconducting gap opens at \tc.  It is apparently independent of structure, inhomogeneity, or "sample quality." 

For most of the doping range, this pseudogap matter, despite the fast and metallic ($1/T_1\propto T$) relaxation rate, has a vanishing uniform response (spin shift), something which is not expected for a simple metal. The temperature dependence of the shifts within the pseudogap regime apparently results from decreased excitations above the gap with decreasing temperature. 

\begin{figure}[h!]
\centering
\includegraphics[width=.6\textwidth ]{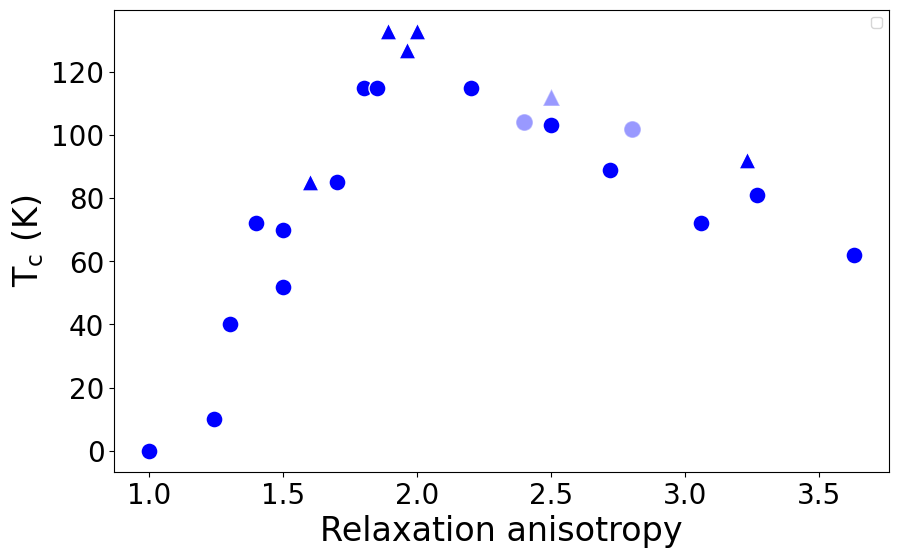}
\caption{Quoted value of \tc plotted against the Cu relaxation anisotropy (${^{63}R}$) of the pseudogap matter for a variety of materials and doping levels. Optimally doped materials within a given family are marked with triangles instead of circles, and the faded data points indicate members of the Tl-2212 family, for which $1/T_{1\perp}$ was not directly measured (see Appendix for details). Increased doping within a family decreases the relaxation anisotropy. Note that members of the \ybco family show significantly larger relaxation anisotropies compared to other materials at similar doping levels, including the optimally doped compound, but nevertheless lie on the same curve. For a list of material-dependent anisotropies, see Table \ref{tab:table1} in the Appendix. The anisotropies plotted here result from the true, unscaled relaxation data, and the estimated error in the anisotropy is $\leq 0.035$.}\label{fig:fig5}
\end{figure}

In the pseudogap matter regime, the Cu relaxation anisotropy is temperature independent, but doping and family dependent. $1/^{63}T_{1\parallel}T$ varies greatly between materials, and is primarily responsible for the different relaxation anisotropies. However, it is proportional to $1/^{63}T_{1\perp}T$ throughout this region of the phase diagram. Most remarkably, the relaxation anisotropy measured on this matter directly encodes \tc, as is shown in Fig. \ref{fig:fig5}. The highest \tc materials have a relaxation anisotropy $^{63}R\approx1.9$. Optimally doped materials with a lower \tcmax lie at either higher or lower anisotropy, depending on the family, but lie on the same lines as the non-optimally doped samples. There is apparently an ideal makeup of the pseudogap matter, from which the superconducting state emerges, which is essential for a high \tc.  

It is thus important to understand what is behind this relaxation anisotropy. Nuclear relaxation probes the low-energy fluctuation spectrum as projected through the hyperfine couplings with their associated form factors, resulting in different relaxation rates for different directions of the applied field. Fluctuating fields perpendicular to the applied field will induce relaxation. Shift analyses \cite{Bandur2026, Lee2026s} show that planar Cu is coupled to two spin components, A and B, through the hyperfine couplings $A_\alpha$ (anisotropic) and $B$ (isotropic), so both spin components should play a role in the relaxation. Meanwhile, planar O couples only to the A-spin through the hyperfine $C_\alpha$. Such analysis further demonstrates that the hyperfine couplings must be rather family and doping independent, such that the variation in the shifts is rooted in the electronic spin components themselves. From planar O shift, we know that the A-spin uniform response at high doping $x$ is family independent \cite{Nachtigal2020}, while Cu tells us that the B-spin uniform response is also family independent, although it shows a strong doping dependence \cite{Lee2026s}. Thus we cannot explain the family dependence of the relaxation anisotropy through the variation in the shift components alone.

We then have to look beyond the shifts. With a negative $A_\parallel$ and positive $A_\perp$ and $B$, the contribution to relaxation from antiferromagnetic A-B fluctuations will appear amplified in the out-of-plane direction from the perspective of the nucleus, whereas in-plane they will appear diminished. An increase in such antiferromagnetic fluctuations will result in a larger anisotropy and vice versa. The anisotropy is therefore expected to provide information regarding the balance of antiferromagnetic vs. more metallic excitations. Thus the family dependence encoded in the relation between $^{63}R$ and \tc presumably stems from the q-dependent fluctuation spectrum. This also generally fits what is known from neutron scattering \cite{Fujita2012}, which shows a doping and family dependent peak near the antiferromagnetic wavevector.

\begin{figure}[h!]
\centering
\includegraphics[width=.6\textwidth ]{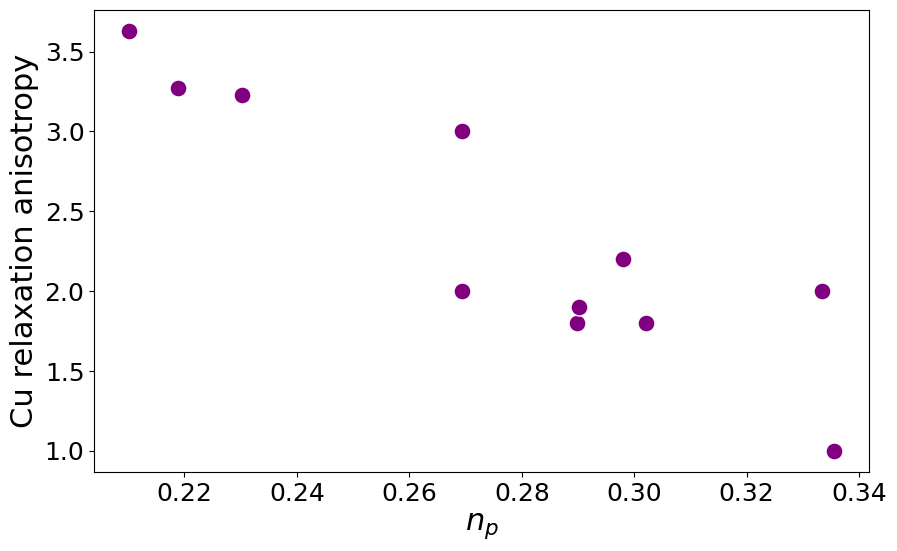}
\caption{Planar Cu relaxation anisotropy (${^{63}R}$) decreases with increasing planar O hole content $n_p$. Planar O quadrupole splittings are taken from \cite{Rybicki2016}, wherein it was also shown that the planar O hole content at optimal doping correlates with \tcmax. Note that this planar O hole content includes both the doped as well as shared parent hole \cite{Jurkutat2014}.}\label{fig:fig6}
\end{figure}

It has previously been shown that the maximum critical temperature of a family is proportional to the total planar O hole content (distributed parent plus doped holes), with $T_{\mathrm{c,max}}\propto 2n_p(x_\mathrm{{opt}})$ \cite{Rybicki2016}. The relationship has also been confirmed with c-DMFT and related to the charge transfer gap \cite{Kowalski2021}. Here, as shown in Fig. \ref{fig:fig6}, we find a correlation between the Cu relaxation anisotropy and the planar O hole content, connecting the planar charge data to the magnetic.

Definitions of $T^*$ from shift data (both that of Cu and O) are rooted in the size of the pseudogap. These show the pseudogap terminating at a doping $p^*\approx0.2$ before the end of the superconducting dome. The shifts, however, are not able to probe the pseudogap matter directly; rather, they probe the states above it and are restricted to give information about the uniform response, whereas the relaxation (with some filter factor applied through the hyperfine constant) probes the low energy fluctuations which can relax the nuclear spin. This filtering is presumably why planar O relaxation apparently sees almost none of the pseudogap matter excitations, and is instead dictated predominantly by excitations above the pseudogap - the hyperfine filtered contributions from the two neighboring Cu atoms should cancel in the case of antiferromagnetic A-spin fluctuations. We see a planar O relaxation that is proportional to $2C^2(1+\rho)$, where $\rho$ is simply the (negative, antiferromagnetic) correlation coefficient between neighboring Cu spins. Thus neither O shift and relaxation nor Cu shift data are able to measure the excitations which are present in the pseudogap regime, only those outside it. Cu relaxation, on the other hand, is able to probe the excitations of the pseudogap matter directly, Such differences are important to keep in mind when considering the controversy over the true endpoint of the pseudogap. Based upon the Cu relaxation, it seems that the pseudogap matter can appear (in the overdoped materials, for example) without a clear gap in the density of states.


We next discuss region III in Fig. \ref{fig:fig4}. In this region, planar O relaxation is proportional to temperature as expected for a metal, but with an offset interpreted as missing low-energy states. It also approximately obeys the Korringa relation \cite{Nachtigal2020} at high enough temperatures. Similarly, the planar Cu  A- and B-shifts also appear metallic in this regime, unlike the relaxation at first glance. However, it turns out that $1/^{63}T_1$ is also linear, but in $T-T^*$ rather than the bare $T$ (see inset to Fig. \ref{fig:fig3}). Upon subtracting the value at $T^*$ (thus shifting the points of departure from the pseudogap matter to the origin), this renormalized metal follows a new Korringa-like scaling (constant $1/T_{1}T$). In this light, the change in the temperature dependence of the planar O relaxation near this temperature may also follow from such a renormalization. 

Interestingly, when we look to this renormalized scaling, we find good agreement to the A- and B-shifts through the Korringa relation if we add their respective associated contributions to the relaxation incoherently, i.e. as $K_A^2+K_B^2$ where the subscript label indicates the associated susceptibilities. For the most highly overdoped (non-superconducting) sample, we get from the shifts a two-component metallic relaxation of 18/Ks, similar to the measured value of 20/Ks of the renormalized metal above $T^*$. Decreasing doping tends to decrease the high-T slopes, consistent with the decreasing shift components. This relaxation contribution sits on top of that of the pseudogap matter, which seems to remain underneath at lower energies, consistent with previous considerations of the pseudogap as temperature independent. Thus above $T^*$ we seem to enter a new, two-component metallic phase with a renormalized temperature scale. Unfortunately, fewer complete data sets are available in this region, particularly at very high temperatures, where the behavior may change \cite{Imai1994}. 

Below $T^*$ on the other hand, we find better agreement between the shifts and relaxation if we allow for some degree of antiferromagnetic coherence between A- and B-spins, as in $(K_A-K_B)^2$. Such a coherence, given the symmetries of the hyperfine couplings and involved form factors, would be expected to increase the anisotopy of the relaxation upon crossing into the pseudogap regime from above, which is what we observe in experimental data. A greater degree of antiferromagnetic correlation (embedded in the cross-term $-2\erww{K_A\cdot K_B}$) should increase relaxation anisotropy, and thus correlations between the A- and B-spins should be an important, albeit complex, control parameter for the anisotropy, and with it \tc.

The phenomenology presented here should certainly manifest itself in observations from other probes as well. For example, STM \cite{Fischer2007} and specific heat \cite{Loram1998, Tallon2022} also show a temperature-independent pseudogap which closes as a function of doping. These are expected to weight the antinodal regions near the saddle point more heavily \cite{Fujita2012b}, as does planar O relaxation, given the form factor, which suppresses fluctuations near the antiferromagnetic wavevector. Transport data, on the other hand, is known to weight the nodal regions more heavily, where $v_F$ is larger and quasiparticle lifetimes are longer, and shows metallic ($\rho\propto T^2$) behavior in the pseudogap regime \cite{Barisic2013}. Considering the theoretical form factors for planar Cu hyperfine coupling \cite{Mila1989b, Chen2017}, this should be the manifestation of the hidden metal we describe here.  The independence of the planar Cu relaxation with \cperp on doping, while the relaxation anisotropy changes, should be tied to the shifting distribution of spectral weight seen by neutron scattering \cite{Fujita2012, Birgeneau2006} and optical conductivity \cite{Kumar2023}. We would also like to note that the two-component phenomenology presented here may invite analogies to Kondo lattice physics.

It is also worth considering that the apparently disparate endpoints of the hidden metal and the thermodynamic pseudogap could also result if the former tracks a different but closely related parameter, such as a doping-dependent saturation of correlation length or magnetic crossover into a correlated metal \cite{Curro1997,Sordi2012, Sordi2012b} than to the actual gap in the normal-state DOS observed by planar O, which is blind to the states that maintain this hidden metal. We would like to note here that the line between regions I and III in Fig. \ref{fig:fig4} would extrapolate at zero doping to a value near $T_N$, but there is not enough data here at low doping to explore this further. 

Finally, we would like to note that the observation of an optimal relaxation anisotropy is similar to the previously observed matching condition between the Cu shift components \cite{Bandur2026}. The matching condition in the Cu shift components seems to determine optimal doping, but not the family-dependent \tcmax that is encoded in the relaxation anisotropy.

In summary, the pseudogap matter is related to this hidden metal, as is best observed in planar Cu nuclear relaxation. A profound implication of the phenomenology is that there seems to be no fundamental distinction between the emergence of superconductivity in the underdoped vs. the overdoped materials; on both ends of the phase diagram, superconductivity set in from this pseudogap matter. The behavior is remarkably universal across all cuprates without any consideration for structural differences, sample quality, or other material-dependent properties. It is not clear at present what role charge and spin inhomogeneities, which have been observed in the NMR data \cite{Haase2002, Rybicki2009}, play in this phenomenology. The role of the lattice \cite{Conradson2020} in the pseudogap matter also remains as an avenue of further discussion. It is known that there is a large isotope effect on $T^*$ \cite{Keller2008}, but not yet how may relate to the renormalization we observe in the relaxation. The pseudogap matter is metallic in terms of the temperature dependence of the Cu relaxation, meaning that $1/^{63}T_1T$ is constant within its region of the phase diagram. However, this pseudogap matter predominantly shows a vanishing uniform response as measured via the spin shifts, in stark contrast to shifts in an ordinary Fermi liquid. This metallic relaxation, along with its associated anisotropy, changes above the pseudogap, developing a lower anisotropy, and a smaller, but still linear temperature dependence, if renormalized appropriately. This regime seems to be much closer to that expected for an ordinary metal, but has two spin components. Given the relationship between \tcmax and the relaxation anisotropy, we conclude that it is the relationship between both spin components, rather than either individually, which ultimately controls superconductivity in the cuprates.


\begin{acknowledgements}
We acknowledge financial support from Leipzig University, in particular Roger Gläser, and stimulating scientific discussions with Nabeel Aslam. We also thank Steve Conradson for valuable feedback. 
\end{acknowledgements}

{\flushleft \bf Author contributions\\}
A.L. performed all data analysis. A.L. and J.H. contributed equally to preparing the manuscript. J.H. had the overall leadership.

{\flushleft \bf Competing Interests\\}
{ There are no competing interests for any of the authors.}\par\medskip

{\flushleft \bf Data Availability Statement \\}
All used data will be made available which enables reproduction of the results.\par\medskip

{\flushleft \bf Funding\\}
Funding of the research came from Leipzig University.\par\medskip
\vspace{0.5cm}

\bibliography{JH-CuprateX.bib}   

\begin{thebibliography}{58}%
\makeatletter
\providecommand \@ifxundefined [1]{%
 \@ifx{#1\undefined}
}%
\providecommand \@ifnum [1]{%
 \ifnum #1\expandafter \@firstoftwo
 \else \expandafter \@secondoftwo
 \fi
}%
\providecommand \@ifx [1]{%
 \ifx #1\expandafter \@firstoftwo
 \else \expandafter \@secondoftwo
 \fi
}%
\providecommand \natexlab [1]{#1}%
\providecommand \enquote  [1]{``#1''}%
\providecommand \bibnamefont  [1]{#1}%
\providecommand \bibfnamefont [1]{#1}%
\providecommand \citenamefont [1]{#1}%
\providecommand \href@noop [0]{\@secondoftwo}%
\providecommand \href [0]{\begingroup \@sanitize@url \@href}%
\providecommand \@href[1]{\@@startlink{#1}\@@href}%
\providecommand \@@href[1]{\endgroup#1\@@endlink}%
\providecommand \@sanitize@url [0]{\catcode `\\12\catcode `\$12\catcode
  `\&12\catcode `\#12\catcode `\^12\catcode `\_12\catcode `\%12\relax}%
\providecommand \@@startlink[1]{}%
\providecommand \@@endlink[0]{}%
\providecommand \url  [0]{\begingroup\@sanitize@url \@url }%
\providecommand \@url [1]{\endgroup\@href {#1}{\urlprefix }}%
\providecommand \urlprefix  [0]{URL }%
\providecommand \Eprint [0]{\href }%
\providecommand \doibase [0]{https://doi.org/}%
\providecommand \selectlanguage [0]{\@gobble}%
\providecommand \bibinfo  [0]{\@secondoftwo}%
\providecommand \bibfield  [0]{\@secondoftwo}%
\providecommand \translation [1]{[#1]}%
\providecommand \BibitemOpen [0]{}%
\providecommand \bibitemStop [0]{}%
\providecommand \bibitemNoStop [0]{.\EOS\space}%
\providecommand \EOS [0]{\spacefactor3000\relax}%
\providecommand \BibitemShut  [1]{\csname bibitem#1\endcsname}%
\let\auto@bib@innerbib\@empty
\bibitem [{\citenamefont {Slichter}(1990)}]{Slichter1990}%
  \BibitemOpen
  \bibfield  {author} {\bibinfo {author} {\bibfnamefont {C.~P.}\ \bibnamefont
  {Slichter}},\ }\href@noop {} {\emph {\bibinfo {title} {{Principles of
  Magnetic Resonance}}}},\ \bibinfo {edition} {third enlarged}\ ed.\ (\bibinfo
  {publisher} {Springer},\ \bibinfo {address} {Berlin},\ \bibinfo {year}
  {1990})\BibitemShut {NoStop}%
\bibitem [{\citenamefont {Heitler}\ and\ \citenamefont
  {Teller}(1936)}]{Heitler1936}%
  \BibitemOpen
  \bibfield  {author} {\bibinfo {author} {\bibfnamefont {W.}~\bibnamefont
  {Heitler}}\ and\ \bibinfo {author} {\bibfnamefont {E.}~\bibnamefont
  {Teller}},\ }\bibfield  {title} {\bibinfo {title} {{Time Effects in the
  Magnetic Cooling Method-I}},\ }\href {https://doi.org/10.1098/rspa.1936.0124}
  {\bibfield  {journal} {\bibinfo  {journal} {Proc. R. Soc. A Math. Phys. Eng.
  Sci.}\ }\textbf {\bibinfo {volume} {155}},\ \bibinfo {pages} {629} (\bibinfo
  {year} {1936})}\BibitemShut {NoStop}%
\bibitem [{\citenamefont {Korringa}(1950)}]{Korringa1950}%
  \BibitemOpen
  \bibfield  {author} {\bibinfo {author} {\bibfnamefont {J.}~\bibnamefont
  {Korringa}},\ }\bibfield  {title} {\bibinfo {title} {{Nuclear magnetic
  relaxation and resonnance line shift in metals}},\ }\href
  {https://doi.org/10.1016/0031-8914(50)90105-4} {\bibfield  {journal}
  {\bibinfo  {journal} {Physica}\ }\textbf {\bibinfo {volume} {16}},\ \bibinfo
  {pages} {601} (\bibinfo {year} {1950})}\BibitemShut {NoStop}%
\bibitem [{\citenamefont {Hebel}\ and\ \citenamefont
  {Slichter}(1959)}]{Hebel1957}%
  \BibitemOpen
  \bibfield  {author} {\bibinfo {author} {\bibfnamefont {L.~C.}\ \bibnamefont
  {Hebel}}\ and\ \bibinfo {author} {\bibfnamefont {C.~P.}\ \bibnamefont
  {Slichter}},\ }\bibfield  {title} {\bibinfo {title} {{Nuclear Spin Relaxation
  in Normal and Superconducting Aluminum}},\ }\href
  {https://doi.org/10.1103/PhysRev.113.1504} {\bibfield  {journal} {\bibinfo
  {journal} {Phys. Rev.}\ }\textbf {\bibinfo {volume} {113}},\ \bibinfo {pages}
  {1504} (\bibinfo {year} {1959})}\BibitemShut {NoStop}%
\bibitem [{\citenamefont {Bednorz}\ and\ \citenamefont
  {M{\"{u}}ller}(1986)}]{Bednorz1986}%
  \BibitemOpen
  \bibfield  {author} {\bibinfo {author} {\bibfnamefont {J.~G.}\ \bibnamefont
  {Bednorz}}\ and\ \bibinfo {author} {\bibfnamefont {K.~A.}\ \bibnamefont
  {M{\"{u}}ller}},\ }\bibfield  {title} {\bibinfo {title} {{Possible High
  $T_\mathrm{c}$ Superconductivity in the Ba-La-Cu-O System}},\ }\href
  {https://doi.org/10.1007/BF01303701} {\bibfield  {journal} {\bibinfo
  {journal} {Z. Phys. B Condens. Matter}\ }\textbf {\bibinfo {volume} {193}},\
  \bibinfo {pages} {189} (\bibinfo {year} {1986})}\BibitemShut {NoStop}%
\bibitem [{\citenamefont {Suter}\ \emph {et~al.}(2000)\citenamefont {Suter},
  \citenamefont {Mali}, \citenamefont {Roos},\ and\ \citenamefont
  {Brinkmann}}]{Suter2000}%
  \BibitemOpen
  \bibfield  {author} {\bibinfo {author} {\bibfnamefont {A.}~\bibnamefont
  {Suter}}, \bibinfo {author} {\bibfnamefont {M.}~\bibnamefont {Mali}},
  \bibinfo {author} {\bibfnamefont {J.}~\bibnamefont {Roos}},\ and\ \bibinfo
  {author} {\bibfnamefont {D.}~\bibnamefont {Brinkmann}},\ }\bibfield  {title}
  {\bibinfo {title} {{Charge degree of freedom and the single-spin fluid model
  in YBa$_2$Cu$_4$O$_8$}},\ }\href
  {https://doi.org/10.1103/PhysRevLett.84.4938} {\bibfield  {journal} {\bibinfo
   {journal} {Phys. Rev. Lett.}\ }\textbf {\bibinfo {volume} {84}},\ \bibinfo
  {pages} {4938} (\bibinfo {year} {2000})}\BibitemShut {NoStop}%
\bibitem [{\citenamefont {Zheng}\ \emph {et~al.}(1995)\citenamefont {Zheng},
  \citenamefont {Kitaoka}, \citenamefont {Ishida},\ and\ \citenamefont
  {Asayama}}]{Zheng1995h}%
  \BibitemOpen
  \bibfield  {author} {\bibinfo {author} {\bibfnamefont {G.-q.}\ \bibnamefont
  {Zheng}}, \bibinfo {author} {\bibfnamefont {Y.}~\bibnamefont {Kitaoka}},
  \bibinfo {author} {\bibfnamefont {K.}~\bibnamefont {Ishida}},\ and\ \bibinfo
  {author} {\bibfnamefont {K.}~\bibnamefont {Asayama}},\ }\bibfield  {title}
  {\bibinfo {title} {Local {Hole} {Distribution} in the {CuO$_2$} {Plane} of
  {High}-{T$_\mathrm{c}$} {Cu}-{Oxides} {Studied} by {Cu} and {Oxygen}
  {NQR}/{NMR}},\ }\href {https://doi.org/10.1143/JPSJ.64.2524} {\bibfield
  {journal} {\bibinfo  {journal} {J. Phys. Soc. Jpn.}\ }\textbf {\bibinfo
  {volume} {64}},\ \bibinfo {pages} {2524} (\bibinfo {year}
  {1995})}\BibitemShut {NoStop}%
\bibitem [{\citenamefont {Jurkutat}\ \emph {et~al.}(2014)\citenamefont
  {Jurkutat}, \citenamefont {Rybicki}, \citenamefont {Sushkov}, \citenamefont
  {Williams}, \citenamefont {Erb},\ and\ \citenamefont {Haase}}]{Jurkutat2014}%
  \BibitemOpen
  \bibfield  {author} {\bibinfo {author} {\bibfnamefont {M.}~\bibnamefont
  {Jurkutat}}, \bibinfo {author} {\bibfnamefont {D.}~\bibnamefont {Rybicki}},
  \bibinfo {author} {\bibfnamefont {O.~P.}\ \bibnamefont {Sushkov}}, \bibinfo
  {author} {\bibfnamefont {G.~V.~M.}\ \bibnamefont {Williams}}, \bibinfo
  {author} {\bibfnamefont {A.}~\bibnamefont {Erb}},\ and\ \bibinfo {author}
  {\bibfnamefont {J.}~\bibnamefont {Haase}},\ }\bibfield  {title} {\bibinfo
  {title} {{Distribution of electrons and holes in cuprate superconductors as
  determined from $^{17}$O and $^{63}$Cu nuclear magnetic resonance}},\ }\href
  {https://doi.org/10.1103/PhysRevB.90.140504} {\bibfield  {journal} {\bibinfo
  {journal} {Phys. Rev. B}\ }\textbf {\bibinfo {volume} {90}},\ \bibinfo
  {pages} {140504(R)} (\bibinfo {year} {2014})}\BibitemShut {NoStop}%
\bibitem [{\citenamefont {Rybicki}\ \emph {et~al.}(2016)\citenamefont
  {Rybicki}, \citenamefont {Jurkutat}, \citenamefont {Reichardt}, \citenamefont
  {Kapusta},\ and\ \citenamefont {Haase}}]{Rybicki2016}%
  \BibitemOpen
  \bibfield  {author} {\bibinfo {author} {\bibfnamefont {D.}~\bibnamefont
  {Rybicki}}, \bibinfo {author} {\bibfnamefont {M.}~\bibnamefont {Jurkutat}},
  \bibinfo {author} {\bibfnamefont {S.}~\bibnamefont {Reichardt}}, \bibinfo
  {author} {\bibfnamefont {C.}~\bibnamefont {Kapusta}},\ and\ \bibinfo {author}
  {\bibfnamefont {J.}~\bibnamefont {Haase}},\ }\bibfield  {title} {\bibinfo
  {title} {{Perspective on the phase diagram of cuprate high-temperature
  superconductors}},\ }\href {https://doi.org/10.1038/ncomms11413} {\bibfield
  {journal} {\bibinfo  {journal} {Nat. Commun.}\ }\textbf {\bibinfo {volume}
  {7}},\ \bibinfo {pages} {11413} (\bibinfo {year} {2016})}\BibitemShut
  {NoStop}%
\bibitem [{\citenamefont {Kowalski}\ \emph {et~al.}(2021)\citenamefont
  {Kowalski}, \citenamefont {Dash}, \citenamefont {S{\'e}mon}, \citenamefont
  {S{\'e}n{\'e}chal},\ and\ \citenamefont {Tremblay}}]{Kowalski2021}%
  \BibitemOpen
  \bibfield  {author} {\bibinfo {author} {\bibfnamefont {N.}~\bibnamefont
  {Kowalski}}, \bibinfo {author} {\bibfnamefont {S.~S.}\ \bibnamefont {Dash}},
  \bibinfo {author} {\bibfnamefont {P.}~\bibnamefont {S{\'e}mon}}, \bibinfo
  {author} {\bibfnamefont {D.}~\bibnamefont {S{\'e}n{\'e}chal}},\ and\ \bibinfo
  {author} {\bibfnamefont {A.-M.}\ \bibnamefont {Tremblay}},\ }\bibfield
  {title} {\bibinfo {title} {Oxygen hole content, charge-transfer gap,
  covalency, and cuprate superconductivity},\ }\href
  {https://doi.org/10.1073/pnas.2106476118} {\bibfield  {journal} {\bibinfo
  {journal} {PNAS}\ }\textbf {\bibinfo {volume} {118}},\ \bibinfo {pages}
  {e2106476118} (\bibinfo {year} {2021})}\BibitemShut {NoStop}%
\bibitem [{\citenamefont {Walstedt}\ \emph {et~al.}(1988)\citenamefont
  {Walstedt}, \citenamefont {W~W~Warren}, \citenamefont {Bell}, \citenamefont
  {Brennert}, \citenamefont {Espinosa}, \citenamefont {Cava}, \citenamefont
  {Schneemeyer},\ and\ \citenamefont {Waszczak}}]{Walstedt1988}%
  \BibitemOpen
  \bibfield  {author} {\bibinfo {author} {\bibfnamefont {R.~E.}\ \bibnamefont
  {Walstedt}}, \bibinfo {author} {\bibfnamefont {J.}~\bibnamefont
  {W~W~Warren}}, \bibinfo {author} {\bibfnamefont {R.~F.}\ \bibnamefont
  {Bell}}, \bibinfo {author} {\bibfnamefont {G.~F.}\ \bibnamefont {Brennert}},
  \bibinfo {author} {\bibfnamefont {G.~P.}\ \bibnamefont {Espinosa}}, \bibinfo
  {author} {\bibfnamefont {R.~J.}\ \bibnamefont {Cava}}, \bibinfo {author}
  {\bibfnamefont {L.~F.}\ \bibnamefont {Schneemeyer}},\ and\ \bibinfo {author}
  {\bibfnamefont {J.~V.}\ \bibnamefont {Waszczak}},\ }\bibfield  {title}
  {\bibinfo {title} {Anisotropic nuclear relaxation in {YBa$_2$Cu$_3$O$_7$}},\
  }\href@noop {} {\bibfield  {journal} {\bibinfo  {journal} {Phys. Rev. B}\
  }\textbf {\bibinfo {volume} {38}},\ \bibinfo {pages} {9299} (\bibinfo {year}
  {1988})}\BibitemShut {NoStop}%
\bibitem [{\citenamefont {Takigawa}\ \emph {et~al.}(1989)\citenamefont
  {Takigawa}, \citenamefont {Hammel}, \citenamefont {Heffner},\ and\
  \citenamefont {Fisk}}]{Takigawa1989}%
  \BibitemOpen
  \bibfield  {author} {\bibinfo {author} {\bibfnamefont {M.}~\bibnamefont
  {Takigawa}}, \bibinfo {author} {\bibfnamefont {P.~C.}\ \bibnamefont
  {Hammel}}, \bibinfo {author} {\bibfnamefont {R.~H.}\ \bibnamefont
  {Heffner}},\ and\ \bibinfo {author} {\bibfnamefont {Z.}~\bibnamefont
  {Fisk}},\ }\bibfield  {title} {\bibinfo {title} {{Spin susceptibility in
  superconducting YBa$_{2}$Cu$_{3}$O$_{7}$ from $^{63}$Cu Knight shift}},\
  }\href {https://doi.org/10.1103/PhysRevB.39.7371} {\bibfield  {journal}
  {\bibinfo  {journal} {Phys. Rev. B}\ }\textbf {\bibinfo {volume} {39}},\
  \bibinfo {pages} {7371} (\bibinfo {year} {1989})}\BibitemShut {NoStop}%
\bibitem [{\citenamefont {Pennington}\ \emph {et~al.}(1989)\citenamefont
  {Pennington}, \citenamefont {Durand}, \citenamefont {Slichter}, \citenamefont
  {Rice}, \citenamefont {Bukowski},\ and\ \citenamefont
  {Ginsberg}}]{Pennington1989}%
  \BibitemOpen
  \bibfield  {author} {\bibinfo {author} {\bibfnamefont {C.~H.}\ \bibnamefont
  {Pennington}}, \bibinfo {author} {\bibfnamefont {D.~J.}\ \bibnamefont
  {Durand}}, \bibinfo {author} {\bibfnamefont {C.~P.}\ \bibnamefont
  {Slichter}}, \bibinfo {author} {\bibfnamefont {J.~P.}\ \bibnamefont {Rice}},
  \bibinfo {author} {\bibfnamefont {E.~D.}\ \bibnamefont {Bukowski}},\ and\
  \bibinfo {author} {\bibfnamefont {D.~M.}\ \bibnamefont {Ginsberg}},\
  }\bibfield  {title} {\bibinfo {title} {{Static and dynamic Cu NMR tensors of
  YBa$_2$Cu$_3$O$_{7-\delta}$}},\ }\href
  {https://doi.org/10.1103/PhysRevB.39.2902} {\bibfield  {journal} {\bibinfo
  {journal} {Phys. Rev. B}\ }\textbf {\bibinfo {volume} {39}},\ \bibinfo
  {pages} {2902} (\bibinfo {year} {1989})}\BibitemShut {NoStop}%
\bibitem [{\citenamefont {Zimmermann}\ \emph {et~al.}(1989)\citenamefont
  {Zimmermann}, \citenamefont {Mali}, \citenamefont {Brinkmann}, \citenamefont
  {Karpinski}, \citenamefont {Kaldis},\ and\ \citenamefont
  {Rusiecki}}]{Zimmermann1989}%
  \BibitemOpen
  \bibfield  {author} {\bibinfo {author} {\bibfnamefont {H.}~\bibnamefont
  {Zimmermann}}, \bibinfo {author} {\bibfnamefont {M.}~\bibnamefont {Mali}},
  \bibinfo {author} {\bibfnamefont {D.}~\bibnamefont {Brinkmann}}, \bibinfo
  {author} {\bibfnamefont {J.}~\bibnamefont {Karpinski}}, \bibinfo {author}
  {\bibfnamefont {E.}~\bibnamefont {Kaldis}},\ and\ \bibinfo {author}
  {\bibfnamefont {S.}~\bibnamefont {Rusiecki}},\ }\bibfield  {title} {\bibinfo
  {title} {{Copper NQR and NMR in the superconductor YBa$_2$Cu$_4$O$_{8+x}$}},\
  }\href@noop {} {\bibfield  {journal} {\bibinfo  {journal} {Phys. C
  Supercond.}\ }\textbf {\bibinfo {volume} {159}},\ \bibinfo {pages} {681}
  (\bibinfo {year} {1989})}\BibitemShut {NoStop}%
\bibitem [{\citenamefont {Mila}\ and\ \citenamefont {Rice}(1989)}]{Mila1989b}%
  \BibitemOpen
  \bibfield  {author} {\bibinfo {author} {\bibfnamefont {F.}~\bibnamefont
  {Mila}}\ and\ \bibinfo {author} {\bibfnamefont {T.~M.}\ \bibnamefont
  {Rice}},\ }\bibfield  {title} {\bibinfo {title} {{Analysis of magnetic
  resonance experiments in YBa$_2$Cu$_3$O$_7$}},\ }\href
  {https://doi.org/10.1016/0921-4534(89)90286-4} {\bibfield  {journal}
  {\bibinfo  {journal} {Phys. C: Supercond.}\ }\textbf {\bibinfo {volume}
  {157}},\ \bibinfo {pages} {561} (\bibinfo {year} {1989})}\BibitemShut
  {NoStop}%
\bibitem [{\citenamefont {Presland}\ \emph {et~al.}(1991)\citenamefont
  {Presland}, \citenamefont {Tallon}, \citenamefont {Buckley}, \citenamefont
  {Liu},\ and\ \citenamefont {Flower}}]{Presland1991}%
  \BibitemOpen
  \bibfield  {author} {\bibinfo {author} {\bibfnamefont {M.}~\bibnamefont
  {Presland}}, \bibinfo {author} {\bibfnamefont {J.}~\bibnamefont {Tallon}},
  \bibinfo {author} {\bibfnamefont {R.}~\bibnamefont {Buckley}}, \bibinfo
  {author} {\bibfnamefont {R.}~\bibnamefont {Liu}},\ and\ \bibinfo {author}
  {\bibfnamefont {N.}~\bibnamefont {Flower}},\ }\bibfield  {title} {\bibinfo
  {title} {{General trends in oxygen stoichiometry effects on $T_{\mathrm{c}}$
  in Bi and Tl superconductors}},\ }\href
  {https://doi.org/10.1016/0921-4534(91)90700-9} {\bibfield  {journal}
  {\bibinfo  {journal} {Physica C: Superconductivity}\ }\textbf {\bibinfo
  {volume} {176}},\ \bibinfo {pages} {95} (\bibinfo {year} {1991})}\BibitemShut
  {NoStop}%
\bibitem [{\citenamefont {Takigawa}\ \emph {et~al.}(1991)\citenamefont
  {Takigawa}, \citenamefont {Reyes}, \citenamefont {Hammel}, \citenamefont
  {Thompson}, \citenamefont {Heffner}, \citenamefont {Fisk},\ and\
  \citenamefont {Ott}}]{Takigawa1991}%
  \BibitemOpen
  \bibfield  {author} {\bibinfo {author} {\bibfnamefont {M.}~\bibnamefont
  {Takigawa}}, \bibinfo {author} {\bibfnamefont {A.~P.}\ \bibnamefont {Reyes}},
  \bibinfo {author} {\bibfnamefont {P.~C.}\ \bibnamefont {Hammel}}, \bibinfo
  {author} {\bibfnamefont {J.~D.}\ \bibnamefont {Thompson}}, \bibinfo {author}
  {\bibfnamefont {R.~H.}\ \bibnamefont {Heffner}}, \bibinfo {author}
  {\bibfnamefont {Z.}~\bibnamefont {Fisk}},\ and\ \bibinfo {author}
  {\bibfnamefont {K.~C.}\ \bibnamefont {Ott}},\ }\bibfield  {title} {\bibinfo
  {title} {{Cu and O NMR studies of the magnetic properties of
  YBa$_2$Cu$_3$O$_{6.63}$ ($T_\mathrm{c}$={62}{K})}},\ }\href
  {https://doi.org/10.1103/PhysRevB.43.247} {\bibfield  {journal} {\bibinfo
  {journal} {Phys. Rev. B}\ }\textbf {\bibinfo {volume} {43}},\ \bibinfo
  {pages} {247} (\bibinfo {year} {1991})}\BibitemShut {NoStop}%
\bibitem [{\citenamefont {Bankay}\ \emph {et~al.}(1994)\citenamefont {Bankay},
  \citenamefont {Mali}, \citenamefont {Roos},\ and\ \citenamefont
  {Brinkmann}}]{Bankay1994}%
  \BibitemOpen
  \bibfield  {author} {\bibinfo {author} {\bibfnamefont {M.}~\bibnamefont
  {Bankay}}, \bibinfo {author} {\bibfnamefont {M.}~\bibnamefont {Mali}},
  \bibinfo {author} {\bibfnamefont {J.}~\bibnamefont {Roos}},\ and\ \bibinfo
  {author} {\bibfnamefont {D.}~\bibnamefont {Brinkmann}},\ }\bibfield  {title}
  {\bibinfo {title} {{Single-spin fluid, spin gap, and d-wave pairing in
  YBa$_2$Cu$_4$O$_8$: A NMR and NQR study}},\ }\href
  {https://doi.org/10.1103/PhysRevB.50.6416} {\bibfield  {journal} {\bibinfo
  {journal} {Phys. Rev. B}\ }\textbf {\bibinfo {volume} {50}},\ \bibinfo
  {pages} {6416} (\bibinfo {year} {1994})}\BibitemShut {NoStop}%
\bibitem [{\citenamefont {Haase}\ \emph {et~al.}(2009)\citenamefont {Haase},
  \citenamefont {Goh}, \citenamefont {Meissner}, \citenamefont {Alireza},\ and\
  \citenamefont {Rybicki}}]{Haase2009}%
  \BibitemOpen
  \bibfield  {author} {\bibinfo {author} {\bibfnamefont {J.}~\bibnamefont
  {Haase}}, \bibinfo {author} {\bibfnamefont {S.~K.}\ \bibnamefont {Goh}},
  \bibinfo {author} {\bibfnamefont {T.}~\bibnamefont {Meissner}}, \bibinfo
  {author} {\bibfnamefont {P.~L.}\ \bibnamefont {Alireza}},\ and\ \bibinfo
  {author} {\bibfnamefont {D.}~\bibnamefont {Rybicki}},\ }\bibfield  {title}
  {\bibinfo {title} {{High sensitivity nuclear magnetic resonance probe for
  anvil cell pressure experiments}},\ }\href
  {https://doi.org/10.1063/1.3183504} {\bibfield  {journal} {\bibinfo
  {journal} {Rev. Sci. Instrum.}\ }\textbf {\bibinfo {volume} {80}},\ \bibinfo
  {pages} {073905} (\bibinfo {year} {2009})}\BibitemShut {NoStop}%
\bibitem [{\citenamefont {Rybicki}\ \emph {et~al.}(2015)\citenamefont
  {Rybicki}, \citenamefont {Kohlrautz}, \citenamefont {Haase}, \citenamefont
  {Greven}, \citenamefont {Zhao}, \citenamefont {Chan}, \citenamefont {Dorow},\
  and\ \citenamefont {Veit}}]{Rybicki2015}%
  \BibitemOpen
  \bibfield  {author} {\bibinfo {author} {\bibfnamefont {D.}~\bibnamefont
  {Rybicki}}, \bibinfo {author} {\bibfnamefont {J.}~\bibnamefont {Kohlrautz}},
  \bibinfo {author} {\bibfnamefont {J.}~\bibnamefont {Haase}}, \bibinfo
  {author} {\bibfnamefont {M.}~\bibnamefont {Greven}}, \bibinfo {author}
  {\bibfnamefont {X.}~\bibnamefont {Zhao}}, \bibinfo {author} {\bibfnamefont
  {M.~K.}\ \bibnamefont {Chan}}, \bibinfo {author} {\bibfnamefont {C.~J.}\
  \bibnamefont {Dorow}},\ and\ \bibinfo {author} {\bibfnamefont {M.~J.}\
  \bibnamefont {Veit}},\ }\bibfield  {title} {\bibinfo {title} {{Electronic
  spin susceptibilities and superconductivity in HgBa$_2$CuO$_{4+\delta}$ from
  nuclear magnetic resonance}},\ }\href
  {https://doi.org/10.1103/PhysRevB.92.081115} {\bibfield  {journal} {\bibinfo
  {journal} {Phys. Rev. B}\ }\textbf {\bibinfo {volume} {92}},\ \bibinfo
  {pages} {081115(R)} (\bibinfo {year} {2015})}\BibitemShut {NoStop}%
\bibitem [{\citenamefont {Bandur}\ \emph {et~al.}(2026)\citenamefont {Bandur},
  \citenamefont {Lee}, \citenamefont {Nachtigal}, \citenamefont {Tsankov},\
  and\ \citenamefont {Haase}}]{Bandur2026}%
  \BibitemOpen
  \bibfield  {author} {\bibinfo {author} {\bibfnamefont {D.}~\bibnamefont
  {Bandur}}, \bibinfo {author} {\bibfnamefont {A.}~\bibnamefont {Lee}},
  \bibinfo {author} {\bibfnamefont {J.}~\bibnamefont {Nachtigal}}, \bibinfo
  {author} {\bibfnamefont {S.}~\bibnamefont {Tsankov}},\ and\ \bibinfo {author}
  {\bibfnamefont {J.}~\bibnamefont {Haase}},\ }\bibfield  {title} {\bibinfo
  {title} {{Tow-Carrier Description of Cuprate Superconductors from NMR}},\
  }\href {https://doi.org/10.3390/condmat11010005} {\bibfield  {journal}
  {\bibinfo  {journal} {Condens. Matter}\ }\textbf {\bibinfo {volume} {11}},\
  \bibinfo {pages} {5} (\bibinfo {year} {2026})}\BibitemShut {NoStop}%
\bibitem [{\citenamefont {Lee}\ and\ \citenamefont {Haase}(2026)}]{Lee2026s}%
  \BibitemOpen
  \bibfield  {author} {\bibinfo {author} {\bibfnamefont {A.}~\bibnamefont
  {Lee}}\ and\ \bibinfo {author} {\bibfnamefont {J.}~\bibnamefont {Haase}},\
  }\bibfield  {title} {\bibinfo {title} {{Pseudogap and Condensation in Cuprate
  Superconductors from NMR}},\ }\href {https://doi.org/10.3390/condmat11020019}
  {\bibfield  {journal} {\bibinfo  {journal} {Condens. Matter}\ }\textbf
  {\bibinfo {volume} {11}},\ \bibinfo {pages} {19} (\bibinfo {year}
  {2026})}\BibitemShut {NoStop}%
\bibitem [{\citenamefont {Jurkutat}\ \emph {et~al.}(2019)\citenamefont
  {Jurkutat}, \citenamefont {Avramovska}, \citenamefont {Williams},
  \citenamefont {Dernbach}, \citenamefont {Pavi{\'c}evi{\'c}},\ and\
  \citenamefont {Haase}}]{Jurkutat2019}%
  \BibitemOpen
  \bibfield  {author} {\bibinfo {author} {\bibfnamefont {M.}~\bibnamefont
  {Jurkutat}}, \bibinfo {author} {\bibfnamefont {M.}~\bibnamefont
  {Avramovska}}, \bibinfo {author} {\bibfnamefont {G.~V.~M.}\ \bibnamefont
  {Williams}}, \bibinfo {author} {\bibfnamefont {D.}~\bibnamefont {Dernbach}},
  \bibinfo {author} {\bibfnamefont {D.}~\bibnamefont {Pavi{\'c}evi{\'c}}},\
  and\ \bibinfo {author} {\bibfnamefont {J.}~\bibnamefont {Haase}},\ }\bibfield
   {title} {\bibinfo {title} {{Phenomenology of $^{63}$Cu Nuclear Relaxation in
  Cuprate Superconductors}},\ }\href {https://doi.org/10.3390/condmat4030067}
  {\bibfield  {journal} {\bibinfo  {journal} {J. Supercond. Nov. Magn.}\
  }\textbf {\bibinfo {volume} {155}},\ \bibinfo {pages} {629} (\bibinfo {year}
  {2019})}\BibitemShut {NoStop}%
\bibitem [{\citenamefont {Nachtigal}\ \emph {et~al.}(2020)\citenamefont
  {Nachtigal}, \citenamefont {Avramovska}, \citenamefont {Erb}, \citenamefont
  {Pavi{\'c}evi{\'c}}, \citenamefont {Guehne},\ and\ \citenamefont
  {Haase}}]{Nachtigal2020}%
  \BibitemOpen
  \bibfield  {author} {\bibinfo {author} {\bibfnamefont {J.}~\bibnamefont
  {Nachtigal}}, \bibinfo {author} {\bibfnamefont {M.}~\bibnamefont
  {Avramovska}}, \bibinfo {author} {\bibfnamefont {A.}~\bibnamefont {Erb}},
  \bibinfo {author} {\bibfnamefont {D.}~\bibnamefont {Pavi{\'c}evi{\'c}}},
  \bibinfo {author} {\bibfnamefont {R.}~\bibnamefont {Guehne}},\ and\ \bibinfo
  {author} {\bibfnamefont {J.}~\bibnamefont {Haase}},\ }\bibfield  {title}
  {\bibinfo {title} {{Temperature-Independent Cuprate Pseudogap from Planar
  Oxygen NMR}},\ }\href {https://doi.org/10.3390/condmat5040066} {\bibfield
  {journal} {\bibinfo  {journal} {Condens. Matter}\ }\textbf {\bibinfo {volume}
  {5}},\ \bibinfo {pages} {66} (\bibinfo {year} {2020})}\BibitemShut {NoStop}%
\bibitem [{\citenamefont {Schrieffer}(2007)}]{Schrieffer2007}%
  \BibitemOpen
  \bibfield  {author} {\bibinfo {author} {\bibfnamefont {J.~R.}\ \bibnamefont
  {Schrieffer}},\ }\href@noop {} {\emph {\bibinfo {title} {{Handbook of
  High-Temperature Superconductivity}}}}\ (\bibinfo  {publisher} {Springer},\
  \bibinfo {year} {2007})\BibitemShut {NoStop}%
\bibitem [{\citenamefont {Dai}\ \emph {et~al.}(2024)\citenamefont {Dai},
  \citenamefont {Kreisel},\ and\ \citenamefont {Andersen~B}}]{Dai2024}%
  \BibitemOpen
  \bibfield  {author} {\bibinfo {author} {\bibfnamefont {Y.}~\bibnamefont
  {Dai}}, \bibinfo {author} {\bibfnamefont {A.}~\bibnamefont {Kreisel}},\ and\
  \bibinfo {author} {\bibfnamefont {M.}~\bibnamefont {Andersen~B}},\ }\bibfield
   {title} {\bibinfo {title} {{Existence of Hebel-Slichter peak in
  unconventional kagome superconductors}},\ }\href
  {https://doi.org/10.1103/PhysRevB.110.144516} {\bibfield  {journal} {\bibinfo
   {journal} {Phys. Rev. B}\ }\textbf {\bibinfo {volume} {110}},\ \bibinfo
  {pages} {144516} (\bibinfo {year} {2024})}\BibitemShut {NoStop}%
\bibitem [{\citenamefont {Loram}\ \emph {et~al.}(1998)\citenamefont {Loram},
  \citenamefont {Mirza}, \citenamefont {Cooper},\ and\ \citenamefont
  {Tallon}}]{Loram1998}%
  \BibitemOpen
  \bibfield  {author} {\bibinfo {author} {\bibfnamefont {J.~W.}\ \bibnamefont
  {Loram}}, \bibinfo {author} {\bibfnamefont {K.~A.}\ \bibnamefont {Mirza}},
  \bibinfo {author} {\bibfnamefont {J.~R.}\ \bibnamefont {Cooper}},\ and\
  \bibinfo {author} {\bibfnamefont {J.~L.}\ \bibnamefont {Tallon}},\ }\bibfield
   {title} {\bibinfo {title} {{Specific heat evidence on the normal state
  pseudogap}},\ }\href {https://doi.org/10.1016/S0022-3697(98)00180-2}
  {\bibfield  {journal} {\bibinfo  {journal} {J. Phys. Chem. Solids}\ }\textbf
  {\bibinfo {volume} {59}},\ \bibinfo {pages} {2091} (\bibinfo {year}
  {1998})}\BibitemShut {NoStop}%
\bibitem [{\citenamefont {Fujita}\ \emph {et~al.}(2012)\citenamefont {Fujita},
  \citenamefont {Hiraka}, \citenamefont {Matsuda}, \citenamefont {Matsuura},
  \citenamefont {M.~Tranquada}, \citenamefont {Wakimoto}, \citenamefont {Xu},\
  and\ \citenamefont {Yamada}}]{Fujita2012}%
  \BibitemOpen
  \bibfield  {author} {\bibinfo {author} {\bibfnamefont {M.}~\bibnamefont
  {Fujita}}, \bibinfo {author} {\bibfnamefont {H.}~\bibnamefont {Hiraka}},
  \bibinfo {author} {\bibfnamefont {M.}~\bibnamefont {Matsuda}}, \bibinfo
  {author} {\bibfnamefont {M.}~\bibnamefont {Matsuura}}, \bibinfo {author}
  {\bibfnamefont {J.}~\bibnamefont {M.~Tranquada}}, \bibinfo {author}
  {\bibfnamefont {S.}~\bibnamefont {Wakimoto}}, \bibinfo {author}
  {\bibfnamefont {G.}~\bibnamefont {Xu}},\ and\ \bibinfo {author}
  {\bibfnamefont {K.}~\bibnamefont {Yamada}},\ }\bibfield  {title} {\bibinfo
  {title} {{Progress in Neutron Scattering Studies of Spin Excitations in
  High-$T_{\mathrm{c}}$ Cuprates}},\ }\href
  {https://doi.org/10.1143/JPSJ.81.011007} {\bibfield  {journal} {\bibinfo
  {journal} {J. Phys. Soc. Jap.}\ }\textbf {\bibinfo {volume} {81}},\ \bibinfo
  {pages} {011007} (\bibinfo {year} {2012})}\BibitemShut {NoStop}%
\bibitem [{\citenamefont {Imai}\ \emph {et~al.}(1994)\citenamefont {Imai},
  \citenamefont {Slichter}, \citenamefont {Yoshimura}, \citenamefont {Katoh},\
  and\ \citenamefont {Kosuge}}]{Imai1994}%
  \BibitemOpen
  \bibfield  {author} {\bibinfo {author} {\bibfnamefont {T.}~\bibnamefont
  {Imai}}, \bibinfo {author} {\bibfnamefont {C.~P.}\ \bibnamefont {Slichter}},
  \bibinfo {author} {\bibfnamefont {K.}~\bibnamefont {Yoshimura}}, \bibinfo
  {author} {\bibfnamefont {M.}~\bibnamefont {Katoh}},\ and\ \bibinfo {author}
  {\bibfnamefont {K.}~\bibnamefont {Kosuge}},\ }\bibfield  {title} {\bibinfo
  {title} {{High-temperature $^{63,65}$Cu NQR and NMR study of the high
  temperature superconductor La$_{2-x}$Sr$_x$CuO$_4$ ($0\leq x \leq 0.15)$}},\
  }\href {https://doi.org/10.1016/0921-4526(94)90262-3} {\bibfield  {journal}
  {\bibinfo  {journal} {Physica B: Condensed Matter}\ }\textbf {\bibinfo
  {volume} {197}},\ \bibinfo {pages} {601} (\bibinfo {year}
  {1994})}\BibitemShut {NoStop}%
\bibitem [{\citenamefont {{Fischer, \O{}ystein and Kugler, Martin and
  Maggio-Aprile, Ivan and Berthod, Christophe and Renner,
  Christoph}}(2007)}]{Fischer2007}%
  \BibitemOpen
  \bibfield  {author} {\bibinfo {author} {\bibnamefont {{Fischer, \O{}ystein
  and Kugler, Martin and Maggio-Aprile, Ivan and Berthod, Christophe and
  Renner, Christoph}}},\ }\bibfield  {title} {\bibinfo {title} {{Scanning
  tunneling spectroscopy of high-temperature superconductors}},\ }\href
  {https://doi.org/10.1103/RevModPhys.79.353} {\bibfield  {journal} {\bibinfo
  {journal} {Rev. Mod. Phys.}\ }\textbf {\bibinfo {volume} {79}},\ \bibinfo
  {pages} {353} (\bibinfo {year} {2007})}\BibitemShut {NoStop}%
\bibitem [{\citenamefont {Tallon}\ and\ \citenamefont
  {Storey}(2022)}]{Tallon2022}%
  \BibitemOpen
  \bibfield  {author} {\bibinfo {author} {\bibfnamefont {J.~L.}\ \bibnamefont
  {Tallon}}\ and\ \bibinfo {author} {\bibfnamefont {J.~G.}\ \bibnamefont
  {Storey}},\ }\bibfield  {title} {\bibinfo {title} {Thermodynamics of the
  pseudogap in cuprates},\ }\href {https://doi.org/10.3389/fphy.2022.1030616}
  {\bibfield  {journal} {\bibinfo  {journal} {Front. Phys.}\ }\textbf {\bibinfo
  {volume} {10}},\ \bibinfo {pages} {1030616} (\bibinfo {year}
  {2022})}\BibitemShut {NoStop}%
\bibitem [{\citenamefont {{Kazuhiro Fujita, Andrew R. Schmidt, Eun-Ah Kim,
  Michael J. Lawler, Dung Hai Lee, J. C. Davis, Hiroshi Eisaki, and Shin-ichi
  Uchida}}(2012)}]{Fujita2012b}%
  \BibitemOpen
  \bibfield  {author} {\bibinfo {author} {\bibnamefont {{Kazuhiro Fujita,
  Andrew R. Schmidt, Eun-Ah Kim, Michael J. Lawler, Dung Hai Lee, J. C. Davis,
  Hiroshi Eisaki, and Shin-ichi Uchida}}},\ }\bibfield  {title} {\bibinfo
  {title} {{Spectroscopic Imaging Scanning Tunneling Microscopy Studies of
  Electronic Structure in the Superconducting and Pseudogap Phases of Cuprate
  High-T$_{\mathrm{c}}$ Superconductors}},\ }\href
  {https://doi.org/10.1143/JPSJ.81.011005} {\bibfield  {journal} {\bibinfo
  {journal} {Journal of the Physical Society of Japan}\ }\textbf {\bibinfo
  {volume} {81}},\ \bibinfo {pages} {011005} (\bibinfo {year}
  {2012})}\BibitemShut {NoStop}%
\bibitem [{\citenamefont {Bari{\v s}i{\'c}}\ \emph {et~al.}(2013)\citenamefont
  {Bari{\v s}i{\'c}}, \citenamefont {Badoux}, \citenamefont {Chan},
  \citenamefont {Dorow}, \citenamefont {Tabis}, \citenamefont {Vignolle},
  \citenamefont {Yu}, \citenamefont {B{\'e}ard}, \citenamefont {Zhao},
  \citenamefont {Proust},\ and\ \citenamefont {Li}}]{Barisic2013}%
  \BibitemOpen
  \bibfield  {author} {\bibinfo {author} {\bibfnamefont {N.}~\bibnamefont
  {Bari{\v s}i{\'c}}}, \bibinfo {author} {\bibfnamefont {S.}~\bibnamefont
  {Badoux}}, \bibinfo {author} {\bibfnamefont {M.~K.}\ \bibnamefont {Chan}},
  \bibinfo {author} {\bibfnamefont {C.}~\bibnamefont {Dorow}}, \bibinfo
  {author} {\bibfnamefont {W.}~\bibnamefont {Tabis}}, \bibinfo {author}
  {\bibfnamefont {B.}~\bibnamefont {Vignolle}}, \bibinfo {author}
  {\bibfnamefont {G.}~\bibnamefont {Yu}}, \bibinfo {author} {\bibfnamefont
  {J.}~\bibnamefont {B{\'e}ard}}, \bibinfo {author} {\bibfnamefont
  {X.}~\bibnamefont {Zhao}}, \bibinfo {author} {\bibfnamefont {C.}~\bibnamefont
  {Proust}},\ and\ \bibinfo {author} {\bibfnamefont {Y.}~\bibnamefont {Li}},\
  }\bibfield  {title} {\bibinfo {title} {{Universal quantum oscillations in the
  underdoped cuprate superconductors}},\ }\href
  {https://doi.org/10.1038/nphys2792} {\bibfield  {journal} {\bibinfo
  {journal} {Nat. Phys.}\ }\textbf {\bibinfo {volume} {9}},\ \bibinfo {pages}
  {761} (\bibinfo {year} {2013})}\BibitemShut {NoStop}%
\bibitem [{\citenamefont {Chen}\ \emph {et~al.}(2017)\citenamefont {Chen},
  \citenamefont {LeBlanc},\ and\ \citenamefont {Gull}}]{Chen2017}%
  \BibitemOpen
  \bibfield  {author} {\bibinfo {author} {\bibfnamefont {X.}~\bibnamefont
  {Chen}}, \bibinfo {author} {\bibfnamefont {J.~P.}\ \bibnamefont {LeBlanc}},\
  and\ \bibinfo {author} {\bibfnamefont {E.}~\bibnamefont {Gull}},\ }\bibfield
  {title} {\bibinfo {title} {{Simulation of the NMR response in the pseudogap
  regime of the cuprates.}},\ }\href@noop {} {\bibfield  {journal} {\bibinfo
  {journal} {Nature Communications}\ }\textbf {\bibinfo {volume} {8}},\
  \bibinfo {pages} {14986} (\bibinfo {year} {2017})}\BibitemShut {NoStop}%
\bibitem [{\citenamefont {J.~Birgeneau}\ \emph {et~al.}(2006)\citenamefont
  {J.~Birgeneau}, \citenamefont {Stock}, \citenamefont {M.~Tranquada},\ and\
  \citenamefont {Yamada}}]{Birgeneau2006}%
  \BibitemOpen
  \bibfield  {author} {\bibinfo {author} {\bibfnamefont {R.}~\bibnamefont
  {J.~Birgeneau}}, \bibinfo {author} {\bibfnamefont {C.}~\bibnamefont {Stock}},
  \bibinfo {author} {\bibfnamefont {J.}~\bibnamefont {M.~Tranquada}},\ and\
  \bibinfo {author} {\bibfnamefont {K.}~\bibnamefont {Yamada}},\ }\bibfield
  {title} {\bibinfo {title} {{Magnetic Neutron Scattering in Hole-Doped Cuprate
  Superconductors}},\ }\href {https://doi.org/10.1143/JPSJ.75.111003}
  {\bibfield  {journal} {\bibinfo  {journal} {J. Phys. Soc. Jpn.}\ }\textbf
  {\bibinfo {volume} {75}},\ \bibinfo {pages} {111003} (\bibinfo {year}
  {2006})}\BibitemShut {NoStop}%
\bibitem [{\citenamefont {{Kumar, C. M. N. and Akrap, A. and Homes, C. C. and
  Martino, E. and Klebel-Knobloch, B. and Tabis, W. and Bari\v{s}i\'{c}, O. S.
  and Sunko, D. K. and Bari\v{s}i\'{c}, N.}}(2023)}]{Kumar2023}%
  \BibitemOpen
  \bibfield  {author} {\bibinfo {author} {\bibnamefont {{Kumar, C. M. N. and
  Akrap, A. and Homes, C. C. and Martino, E. and Klebel-Knobloch, B. and Tabis,
  W. and Bari\v{s}i\'{c}, O. S. and Sunko, D. K. and Bari\v{s}i\'{c}, N.}}},\
  }\bibfield  {title} {\bibinfo {title} {{Characterization of two electronic
  subsystems in cuprates through optical conductivity}},\ }\href
  {https://doi.org/10.1103/PhysRevB.107.144515} {\bibfield  {journal} {\bibinfo
   {journal} {Phys. Rev. B}\ }\textbf {\bibinfo {volume} {107}},\ \bibinfo
  {pages} {144515} (\bibinfo {year} {2023})}\BibitemShut {NoStop}%
\bibitem [{\citenamefont {Curro}\ \emph {et~al.}(1997)\citenamefont {Curro},
  \citenamefont {Imai}, \citenamefont {Slichter},\ and\ \citenamefont
  {Dabrowski}}]{Curro1997}%
  \BibitemOpen
  \bibfield  {author} {\bibinfo {author} {\bibfnamefont {N.~J.}\ \bibnamefont
  {Curro}}, \bibinfo {author} {\bibfnamefont {T.}~\bibnamefont {Imai}},
  \bibinfo {author} {\bibfnamefont {C.~P.}\ \bibnamefont {Slichter}},\ and\
  \bibinfo {author} {\bibfnamefont {B.}~\bibnamefont {Dabrowski}},\ }\bibfield
  {title} {\bibinfo {title} {{High-temperature $^{63}\mathrm{Cu}(2)$ nuclear
  quadrupole and magnetic resonance measurements of
  ${\mathrm{YBa}}_{2}{\mathrm{Cu}}_{4}{\mathrm{O}}_{8}$}},\ }\href
  {https://doi.org/10.1103/PhysRevB.56.877} {\bibfield  {journal} {\bibinfo
  {journal} {Phys. Rev. B}\ }\textbf {\bibinfo {volume} {56}},\ \bibinfo
  {pages} {877} (\bibinfo {year} {1997})}\BibitemShut {NoStop}%
\bibitem [{\citenamefont {Sordi}\ \emph {et~al.}(2012)\citenamefont {Sordi},
  \citenamefont {S\'emon}, \citenamefont {Haule},\ and\ \citenamefont
  {Tremblay}}]{Sordi2012}%
  \BibitemOpen
  \bibfield  {author} {\bibinfo {author} {\bibfnamefont {G.}~\bibnamefont
  {Sordi}}, \bibinfo {author} {\bibfnamefont {P.}~\bibnamefont {S\'emon}},
  \bibinfo {author} {\bibfnamefont {K.}~\bibnamefont {Haule}},\ and\ \bibinfo
  {author} {\bibfnamefont {A.-M.~S.}\ \bibnamefont {Tremblay}},\ }\bibfield
  {title} {\bibinfo {title} {Strong coupling superconductivity, pseudogap, and
  mott transition},\ }\href {https://doi.org/10.1103/PhysRevLett.108.216401}
  {\bibfield  {journal} {\bibinfo  {journal} {Phys. Rev. Lett.}\ }\textbf
  {\bibinfo {volume} {108}},\ \bibinfo {pages} {216401} (\bibinfo {year}
  {2012})}\BibitemShut {NoStop}%
\bibitem [{\citenamefont {{Sordi, G. and S{\'e}mon, P. and Haule, K. and
  Tremblay, A.-M. S.}}(2012)}]{Sordi2012b}%
  \BibitemOpen
  \bibfield  {author} {\bibinfo {author} {\bibnamefont {{Sordi, G. and
  S{\'e}mon, P. and Haule, K. and Tremblay, A.-M. S.}}},\ }\bibfield  {title}
  {\bibinfo {title} {{Pseudogap temperature as a Widom line in doped Mott
  insulators}},\ }\href {https://doi.org/10.1038/srep00547} {\bibfield
  {journal} {\bibinfo  {journal} {Scientific Reports}\ }\textbf {\bibinfo
  {volume} {2}},\ \bibinfo {pages} {547} (\bibinfo {year} {2012})}\BibitemShut
  {NoStop}%
\bibitem [{\citenamefont {Haase}\ \emph {et~al.}(2002)\citenamefont {Haase},
  \citenamefont {Slichter},\ and\ \citenamefont {Milling}}]{Haase2002}%
  \BibitemOpen
  \bibfield  {author} {\bibinfo {author} {\bibfnamefont {J.}~\bibnamefont
  {Haase}}, \bibinfo {author} {\bibfnamefont {C.~P.}\ \bibnamefont
  {Slichter}},\ and\ \bibinfo {author} {\bibfnamefont {C.~T.}\ \bibnamefont
  {Milling}},\ }\bibfield  {title} {\bibinfo {title} {{Static Charge and Spin
  Inhomogeneity in La$_{2-x}$Sr$_x$CuO$_4$ by NMR}},\ }\href
  {https://doi.org/10.1023/A:1021014028677} {\bibfield  {journal} {\bibinfo
  {journal} {J. Supercond.}\ }\textbf {\bibinfo {volume} {15}},\ \bibinfo
  {pages} {339} (\bibinfo {year} {2002})}\BibitemShut {NoStop}%
\bibitem [{\citenamefont {Rybicki}\ \emph {et~al.}(2009)\citenamefont
  {Rybicki}, \citenamefont {Haase}, \citenamefont {Greven}, \citenamefont {Yu},
  \citenamefont {Li}, \citenamefont {Cho},\ and\ \citenamefont
  {Zhao}}]{Rybicki2009}%
  \BibitemOpen
  \bibfield  {author} {\bibinfo {author} {\bibfnamefont {D.}~\bibnamefont
  {Rybicki}}, \bibinfo {author} {\bibfnamefont {J.}~\bibnamefont {Haase}},
  \bibinfo {author} {\bibfnamefont {M.}~\bibnamefont {Greven}}, \bibinfo
  {author} {\bibfnamefont {G.}~\bibnamefont {Yu}}, \bibinfo {author}
  {\bibfnamefont {Y.}~\bibnamefont {Li}}, \bibinfo {author} {\bibfnamefont
  {Y.}~\bibnamefont {Cho}},\ and\ \bibinfo {author} {\bibfnamefont
  {X.}~\bibnamefont {Zhao}},\ }\bibfield  {title} {\bibinfo {title} {{Spatial
  Inhomogeneities in Single-Crystal HgBa$_2$CuO$_{4+\delta}$ from ${}^{63}$Cu
  NMR Spin and Quadrupole Shifts}},\ }\href
  {https://doi.org/10.1007/s10948-008-0376-2} {\bibfield  {journal} {\bibinfo
  {journal} {J. Supercond. Nov. Magn.}\ }\textbf {\bibinfo {volume} {22}},\
  \bibinfo {pages} {179} (\bibinfo {year} {2009})}\BibitemShut {NoStop}%
\bibitem [{\citenamefont {Conradson}\ \emph {et~al.}(2022)\citenamefont
  {Conradson}, \citenamefont {Geballe}, \citenamefont {Jin}, \citenamefont
  {Cao}, \citenamefont {Gauzzi}, \citenamefont {Karppinien}, \citenamefont
  {Baldiozzi}, \citenamefont {Li}, \citenamefont {Gillolo}, \citenamefont
  {Jiang}, \citenamefont {Latimer}, \citenamefont {Mueller},\ and\
  \citenamefont {V}}]{Conradson2020}%
  \BibitemOpen
  \bibfield  {author} {\bibinfo {author} {\bibfnamefont {S.~D.}\ \bibnamefont
  {Conradson}}, \bibinfo {author} {\bibfnamefont {T.~H.}\ \bibnamefont
  {Geballe}}, \bibinfo {author} {\bibfnamefont {C.-Q.}\ \bibnamefont {Jin}},
  \bibinfo {author} {\bibfnamefont {L.-P.}\ \bibnamefont {Cao}}, \bibinfo
  {author} {\bibfnamefont {A.}~\bibnamefont {Gauzzi}}, \bibinfo {author}
  {\bibfnamefont {M.}~\bibnamefont {Karppinien}}, \bibinfo {author}
  {\bibfnamefont {G.}~\bibnamefont {Baldiozzi}}, \bibinfo {author}
  {\bibfnamefont {W.-M.}\ \bibnamefont {Li}}, \bibinfo {author} {\bibfnamefont
  {E.}~\bibnamefont {Gillolo}}, \bibinfo {author} {\bibfnamefont {J.~M.}\
  \bibnamefont {Jiang}}, \bibinfo {author} {\bibfnamefont {M.}~\bibnamefont
  {Latimer}}, \bibinfo {author} {\bibfnamefont {O.}~\bibnamefont {Mueller}},\
  and\ \bibinfo {author} {\bibfnamefont {N.}~\bibnamefont {V}},\ }\bibfield
  {title} {\bibinfo {title} {{Nonadiabatic coupling of the dynamical structure
  to the superconductivity in YSr$_2$Cu$_{2.75}$Mo$_{0.25}$O$_{7.54}$ and
  Sr$_2$CuO$_{3.3}$}},\ }\href {https://doi.org/10.1073/pnas.2018336117}
  {\bibfield  {journal} {\bibinfo  {journal} {PNAS}\ }\textbf {\bibinfo
  {volume} {117}},\ \bibinfo {pages} {33099} (\bibinfo {year}
  {2022})}\BibitemShut {NoStop}%
\bibitem [{\citenamefont {Keller}\ \emph {et~al.}(2008)\citenamefont {Keller},
  \citenamefont {Bussmann-Holder},\ and\ \citenamefont
  {M{\"u}ller}}]{Keller2008}%
  \BibitemOpen
  \bibfield  {author} {\bibinfo {author} {\bibfnamefont {H.}~\bibnamefont
  {Keller}}, \bibinfo {author} {\bibfnamefont {A.}~\bibnamefont
  {Bussmann-Holder}},\ and\ \bibinfo {author} {\bibfnamefont {K.~A.}\
  \bibnamefont {M{\"u}ller}},\ }\bibfield  {title} {\bibinfo {title}
  {{Jahn-Teller physics and high-$T_c$ superconductivity}},\ }\href
  {https://doi.org/10.1016/S1369-7021(08)70178-0} {\bibfield  {journal}
  {\bibinfo  {journal} {Materials Today}\ }\textbf {\bibinfo {volume} {11}},\
  \bibinfo {pages} {38} (\bibinfo {year} {2008})}\BibitemShut {NoStop}%
\bibitem [{\citenamefont {Gippius}\ \emph {et~al.}(1999)\citenamefont
  {Gippius}, \citenamefont {Antipov}, \citenamefont {Hoffmann}, \citenamefont
  {L{\"{u}}ders},\ and\ \citenamefont {Buntkowsky}}]{Gippius1999}%
  \BibitemOpen
  \bibfield  {author} {\bibinfo {author} {\bibfnamefont {A.~A.}\ \bibnamefont
  {Gippius}}, \bibinfo {author} {\bibfnamefont {E.~V.}\ \bibnamefont
  {Antipov}}, \bibinfo {author} {\bibfnamefont {W.}~\bibnamefont {Hoffmann}},
  \bibinfo {author} {\bibfnamefont {K.}~\bibnamefont {L{\"{u}}ders}},\ and\
  \bibinfo {author} {\bibfnamefont {G.}~\bibnamefont {Buntkowsky}},\ }\bibfield
   {title} {\bibinfo {title} {{Low-frequency spin dynamics as probed by
  $^{63}$Cu and $^{199}$Hg NMR in HgBa$_{2}$CuO$_{4+ \delta}$ superconductors
  with different oxygen content}},\ }\href@noop {} {\bibfield  {journal}
  {\bibinfo  {journal} {Phys. Rev. B}\ }\textbf {\bibinfo {volume} {59}},\
  \bibinfo {pages} {654} (\bibinfo {year} {1999})}\BibitemShut {NoStop}%
\bibitem [{\citenamefont {Bobroff}\ \emph {et~al.}(1997)\citenamefont
  {Bobroff}, \citenamefont {Alloul}, \citenamefont {Mendels}, \citenamefont
  {Viallet}, \citenamefont {Marucco},\ and\ \citenamefont
  {Colson}}]{Bobroff1997}%
  \BibitemOpen
  \bibfield  {author} {\bibinfo {author} {\bibfnamefont {J.}~\bibnamefont
  {Bobroff}}, \bibinfo {author} {\bibfnamefont {H.}~\bibnamefont {Alloul}},
  \bibinfo {author} {\bibfnamefont {P.}~\bibnamefont {Mendels}}, \bibinfo
  {author} {\bibfnamefont {V.}~\bibnamefont {Viallet}}, \bibinfo {author}
  {\bibfnamefont {J.~F.}\ \bibnamefont {Marucco}},\ and\ \bibinfo {author}
  {\bibfnamefont {D.}~\bibnamefont {Colson}},\ }\bibfield  {title} {\bibinfo
  {title} {{$^{17}$O NMR Evidence for a Pseudogap in the Monolayer
  HgBa$_{2}$CuO$_{4+ \delta }$}},\ }\href
  {https://doi.org/10.1103/PhysRevLett.78.3757} {\bibfield  {journal} {\bibinfo
   {journal} {Phys. Rev. Lett.}\ }\textbf {\bibinfo {volume} {78}},\ \bibinfo
  {pages} {3757} (\bibinfo {year} {1997})}\BibitemShut {NoStop}%
\bibitem [{\citenamefont {Kambe}\ \emph {et~al.}(1993)\citenamefont {Kambe},
  \citenamefont {Yasuoka}, \citenamefont {Hayashi},\ and\ \citenamefont
  {Ueda}}]{Kambe1993}%
  \BibitemOpen
  \bibfield  {author} {\bibinfo {author} {\bibfnamefont {S.}~\bibnamefont
  {Kambe}}, \bibinfo {author} {\bibfnamefont {H.}~\bibnamefont {Yasuoka}},
  \bibinfo {author} {\bibfnamefont {A.}~\bibnamefont {Hayashi}},\ and\ \bibinfo
  {author} {\bibfnamefont {Y.}~\bibnamefont {Ueda}},\ }\bibfield  {title}
  {\bibinfo {title} {{NMR study of the spin dynamics in Tl$_2$Ba$_2$CuO$_y$
  ($T_\mathrm{c}$=85 K)}},\ }\href {https://doi.org/10.1103/PhysRevB.47.2825}
  {\bibfield  {journal} {\bibinfo  {journal} {Phys. Rev. B}\ }\textbf {\bibinfo
  {volume} {47}},\ \bibinfo {pages} {2825} (\bibinfo {year}
  {1993})}\BibitemShut {NoStop}%
\bibitem [{\citenamefont {Fujiwara}\ \emph {et~al.}(1991)\citenamefont
  {Fujiwara}, \citenamefont {Kitaoka}, \citenamefont {Ishida}, \citenamefont
  {Asayama}, \citenamefont {Shimakawa}, \citenamefont {Manako},\ and\
  \citenamefont {Kubo}}]{Fujiwara1991}%
  \BibitemOpen
  \bibfield  {author} {\bibinfo {author} {\bibfnamefont {K.}~\bibnamefont
  {Fujiwara}}, \bibinfo {author} {\bibfnamefont {Y.}~\bibnamefont {Kitaoka}},
  \bibinfo {author} {\bibfnamefont {K.}~\bibnamefont {Ishida}}, \bibinfo
  {author} {\bibfnamefont {K.}~\bibnamefont {Asayama}}, \bibinfo {author}
  {\bibfnamefont {Y.}~\bibnamefont {Shimakawa}}, \bibinfo {author}
  {\bibfnamefont {T.}~\bibnamefont {Manako}},\ and\ \bibinfo {author}
  {\bibfnamefont {Y.}~\bibnamefont {Kubo}},\ }\bibfield  {title} {\bibinfo
  {title} {{{NMR} and {NQR} studies of superconductivity in heavily doped
  {Tl}$_{\mathrm{2}}${Ba}$_{\mathrm{2}}${CuO}$_{\mathrm{6+y}}$ with a single
  {CuO}$_{\mathrm{2}}$ plane}},\ }\href
  {https://doi.org/10.1016/0921-4534(91)90385-C} {\bibfield  {journal}
  {\bibinfo  {journal} {Physica C: Superconductivity}\ }\textbf {\bibinfo
  {volume} {184}},\ \bibinfo {pages} {207} (\bibinfo {year}
  {1991})}\BibitemShut {NoStop}%
\bibitem [{\citenamefont {Kambe}\ \emph {et~al.}(1991)\citenamefont {Kambe},
  \citenamefont {Yoshinari}, \citenamefont {Yasuoka}, \citenamefont {Hayashi},\
  and\ \citenamefont {Ueda}}]{Kambe1991}%
  \BibitemOpen
  \bibfield  {author} {\bibinfo {author} {\bibfnamefont {S.}~\bibnamefont
  {Kambe}}, \bibinfo {author} {\bibfnamefont {Y.}~\bibnamefont {Yoshinari}},
  \bibinfo {author} {\bibfnamefont {H.}~\bibnamefont {Yasuoka}}, \bibinfo
  {author} {\bibfnamefont {A.}~\bibnamefont {Hayashi}},\ and\ \bibinfo {author}
  {\bibfnamefont {Y.}~\bibnamefont {Ueda}},\ }\bibfield  {title} {\bibinfo
  {title} {{${17}$O, $^{63}$Cu and $^{205}$Tl NMR study of over-doped
  Tl$_2$Ba$_2$CuO$_y$}},\ }\href {https://doi.org/10.1016/0921-4534(91)91814-K}
  {\bibfield  {journal} {\bibinfo  {journal} {Phys. C: Supercond.}\ }\textbf
  {\bibinfo {volume} {185}},\ \bibinfo {pages} {1181} (\bibinfo {year}
  {1991})}\BibitemShut {NoStop}%
\bibitem [{\citenamefont {Magishi}\ \emph {et~al.}(1996)\citenamefont
  {Magishi}, \citenamefont {Kitaoka}, \citenamefont {Zheng}, \citenamefont
  {Asayama}, \citenamefont {Kondo}, \citenamefont {Shimakawa}, \citenamefont
  {Manako},\ and\ \citenamefont {Kubo}}]{Magishi1996}%
  \BibitemOpen
  \bibfield  {author} {\bibinfo {author} {\bibfnamefont {K.}~\bibnamefont
  {Magishi}}, \bibinfo {author} {\bibfnamefont {Y.}~\bibnamefont {Kitaoka}},
  \bibinfo {author} {\bibfnamefont {G.~q.}\ \bibnamefont {Zheng}}, \bibinfo
  {author} {\bibfnamefont {K.}~\bibnamefont {Asayama}}, \bibinfo {author}
  {\bibfnamefont {T.}~\bibnamefont {Kondo}}, \bibinfo {author} {\bibfnamefont
  {Y.}~\bibnamefont {Shimakawa}}, \bibinfo {author} {\bibfnamefont
  {T.}~\bibnamefont {Manako}},\ and\ \bibinfo {author} {\bibfnamefont
  {Y.}~\bibnamefont {Kubo}},\ }\bibfield  {title} {\bibinfo {title} {{Magnetic
  excitation and superconductivity in overdoped TlSr$_2$CaCu$_2$O$_{7-\delta}$:
  A $^{63}$Cu NMR study}},\ }\href {https://doi.org/10.1103/PhysRevB.54.10131}
  {\bibfield  {journal} {\bibinfo  {journal} {Phys. Rev. B}\ }\textbf {\bibinfo
  {volume} {54}},\ \bibinfo {pages} {10131} (\bibinfo {year}
  {1996})}\BibitemShut {NoStop}%
\bibitem [{\citenamefont {Auler}\ \emph {et~al.}(1999)\citenamefont {Auler},
  \citenamefont {Horvati{\'c}}, \citenamefont {Gillet}, \citenamefont
  {Berthier}, \citenamefont {Berthier}, \citenamefont {Carretta}, \citenamefont
  {Kitaoka}, \citenamefont {S{\'e}gransan},\ and\ \citenamefont
  {Henry}}]{Auler1999}%
  \BibitemOpen
  \bibfield  {author} {\bibinfo {author} {\bibfnamefont {T.}~\bibnamefont
  {Auler}}, \bibinfo {author} {\bibfnamefont {M.}~\bibnamefont {Horvati{\'c}}},
  \bibinfo {author} {\bibfnamefont {J.~A.}\ \bibnamefont {Gillet}}, \bibinfo
  {author} {\bibfnamefont {C.}~\bibnamefont {Berthier}}, \bibinfo {author}
  {\bibfnamefont {Y.}~\bibnamefont {Berthier}}, \bibinfo {author}
  {\bibfnamefont {P.}~\bibnamefont {Carretta}}, \bibinfo {author}
  {\bibfnamefont {Y.}~\bibnamefont {Kitaoka}}, \bibinfo {author} {\bibfnamefont
  {P.}~\bibnamefont {S{\'e}gransan}},\ and\ \bibinfo {author} {\bibfnamefont
  {J.~Y.}\ \bibnamefont {Henry}},\ }\bibfield  {title} {\bibinfo {title}
  {{Normal state spin susceptibility in YBa$_2$Cu$_3$O$_{6.92}$ single crystal
  from $^{63}$Cu and $^{89}$Y nuclear magnetic resonance}},\ }\href@noop {}
  {\bibfield  {journal} {\bibinfo  {journal} {Phys. C: Supercond.}\ }\textbf
  {\bibinfo {volume} {313}},\ \bibinfo {pages} {255} (\bibinfo {year}
  {1999})}\BibitemShut {NoStop}%
\bibitem [{\citenamefont {Yoshinari}\ \emph {et~al.}(1992)\citenamefont
  {Yoshinari}, \citenamefont {Yasuoka},\ and\ \citenamefont
  {Ueda}}]{Yoshinari1992}%
  \BibitemOpen
  \bibfield  {author} {\bibinfo {author} {\bibfnamefont {Y.}~\bibnamefont
  {Yoshinari}}, \bibinfo {author} {\bibfnamefont {H.}~\bibnamefont {Yasuoka}},\
  and\ \bibinfo {author} {\bibfnamefont {Y.}~\bibnamefont {Ueda}},\ }\bibfield
  {title} {\bibinfo {title} {{Nuclear Spin Relaxation at Planar Copper and
  Oxygen Sites in YBa$_2$Cu$_3$O$_{6.96}$}},\ }\href
  {https://doi.org/10.1143/JPSJ.61.770} {\bibfield  {journal} {\bibinfo
  {journal} {J. Phys. Soc. Jpn.}\ }\textbf {\bibinfo {volume} {61}},\ \bibinfo
  {pages} {770} (\bibinfo {year} {1992})}\BibitemShut {NoStop}%
\bibitem [{\citenamefont {Walstedt}\ \emph {et~al.}(1989)\citenamefont
  {Walstedt}, \citenamefont {Jr}, \citenamefont {Bell},\ and\ \citenamefont
  {Espinosa}}]{Walstedt1989}%
  \BibitemOpen
  \bibfield  {author} {\bibinfo {author} {\bibfnamefont {R.}~\bibnamefont
  {Walstedt}}, \bibinfo {author} {\bibfnamefont {W.}~\bibnamefont {Jr}},
  \bibinfo {author} {\bibfnamefont {R.}~\bibnamefont {Bell}},\ and\ \bibinfo
  {author} {\bibfnamefont {G.}~\bibnamefont {Espinosa}},\ }\bibfield  {title}
  {\bibinfo {title} {{Anisotropic $^{63}$Cu nuclear relaxation in magnetically
  oriented powdered samples of YBa$_2$Cu$_3$O$_7$.}},\ }\href@noop {}
  {\bibfield  {journal} {\bibinfo  {journal} {Phys. Rev. B}\ }\textbf {\bibinfo
  {volume} {40}},\ \bibinfo {pages} {2572} (\bibinfo {year}
  {1989})}\BibitemShut {NoStop}%
\bibitem [{\citenamefont {Kitaoka}\ \emph {et~al.}(1989)\citenamefont
  {Kitaoka}, \citenamefont {Ishida}, \citenamefont {Fujiwara}, \citenamefont
  {Kondo}, \citenamefont {Asayama}, \citenamefont {Horvatic}, \citenamefont
  {Berthier}, \citenamefont {Butaud}, \citenamefont {Segransan}, \citenamefont
  {Berthier}, \citenamefont {Katayama-Yoshida}, \citenamefont {Okabe},\ and\
  \citenamefont {Takahashi}}]{Kitaoka1989}%
  \BibitemOpen
  \bibfield  {author} {\bibinfo {author} {\bibfnamefont {Y.}~\bibnamefont
  {Kitaoka}}, \bibinfo {author} {\bibfnamefont {K.}~\bibnamefont {Ishida}},
  \bibinfo {author} {\bibfnamefont {F.}~\bibnamefont {Fujiwara}}, \bibinfo
  {author} {\bibfnamefont {T.}~\bibnamefont {Kondo}}, \bibinfo {author}
  {\bibfnamefont {K.}~\bibnamefont {Asayama}}, \bibinfo {author} {\bibfnamefont
  {M.}~\bibnamefont {Horvatic}}, \bibinfo {author} {\bibfnamefont
  {Y.}~\bibnamefont {Berthier}}, \bibinfo {author} {\bibfnamefont
  {P.}~\bibnamefont {Butaud}}, \bibinfo {author} {\bibfnamefont
  {P.}~\bibnamefont {Segransan}}, \bibinfo {author} {\bibfnamefont
  {C.}~\bibnamefont {Berthier}}, \bibinfo {author} {\bibfnamefont
  {H.}~\bibnamefont {Katayama-Yoshida}}, \bibinfo {author} {\bibfnamefont
  {Y.}~\bibnamefont {Okabe}},\ and\ \bibinfo {author} {\bibfnamefont
  {T.}~\bibnamefont {Takahashi}},\ }\href
  {https://doi.org/10.1007/978-3-642-83836-1_26} {\emph {\bibinfo {title}
  {Strong {Correlation} and {Superconductivity}}}},\ edited by\ \bibinfo
  {editor} {\bibfnamefont {K.~A.}\ \bibnamefont {M{\"u}ller}}, \bibinfo
  {editor} {\bibfnamefont {M.}~\bibnamefont {Cardona}}, \bibinfo {editor}
  {\bibfnamefont {P.}~\bibnamefont {Fulde}}, \bibinfo {editor} {\bibfnamefont
  {K.}~\bibnamefont {Von~Klitzing}}, \bibinfo {editor} {\bibfnamefont {H.-J.}\
  \bibnamefont {Queisser}}, \bibinfo {editor} {\bibfnamefont {H.~K.~V.}\
  \bibnamefont {Lotsch}}, \bibinfo {editor} {\bibfnamefont {H.}~\bibnamefont
  {Fukuyama}}, \bibinfo {editor} {\bibfnamefont {S.}~\bibnamefont {Maekawa}},\
  and\ \bibinfo {editor} {\bibfnamefont {A.~P.}\ \bibnamefont {Malozemoff}},\
  Vol.~\bibinfo {volume} {89}\ (\bibinfo  {publisher} {Springer Berlin
  Heidelberg},\ \bibinfo {address} {Berlin, Heidelberg},\ \bibinfo {year}
  {1989})\ pp.\ \bibinfo {pages} {262--273}\BibitemShut {NoStop}%
\bibitem [{\citenamefont {Zimmermann}\ \emph {et~al.}(1991)\citenamefont
  {Zimmermann}, \citenamefont {Mali}, \citenamefont {Bankay},\ and\
  \citenamefont {Brinkmann}}]{Zimmermann1991}%
  \BibitemOpen
  \bibfield  {author} {\bibinfo {author} {\bibfnamefont {H.}~\bibnamefont
  {Zimmermann}}, \bibinfo {author} {\bibfnamefont {M.}~\bibnamefont {Mali}},
  \bibinfo {author} {\bibfnamefont {M.}~\bibnamefont {Bankay}},\ and\ \bibinfo
  {author} {\bibfnamefont {D.}~\bibnamefont {Brinkmann}},\ }\bibfield  {title}
  {\bibinfo {title} {{Anisotropy of $^{63}$Cu Knight shift and spin-lattice
  relaxation in YBa$_2$Cu$_4$O$_8$}},\ }\href@noop {} {\bibfield  {journal}
  {\bibinfo  {journal} {Physica C: Supercond.}\ }\textbf {\bibinfo {volume}
  {185-189}},\ \bibinfo {pages} {1145} (\bibinfo {year} {1991})}\BibitemShut
  {NoStop}%
\bibitem [{\citenamefont {Gerashchenko}\ \emph {et~al.}(1999)\citenamefont
  {Gerashchenko}, \citenamefont {Piskunov}, \citenamefont {Mikhalev},
  \citenamefont {Ananyev}, \citenamefont {Okuluva}, \citenamefont
  {Verkhovskii}, \citenamefont {Yakubovskii}, \citenamefont {Shustov},\ and\
  \citenamefont {Trokiner}}]{Gerashchenko1999}%
  \BibitemOpen
  \bibfield  {author} {\bibinfo {author} {\bibfnamefont {A.~P.}\ \bibnamefont
  {Gerashchenko}}, \bibinfo {author} {\bibfnamefont {Y.~V.}\ \bibnamefont
  {Piskunov}}, \bibinfo {author} {\bibfnamefont {K.}~\bibnamefont {Mikhalev}},
  \bibinfo {author} {\bibfnamefont {A.}~\bibnamefont {Ananyev}}, \bibinfo
  {author} {\bibfnamefont {K.}~\bibnamefont {Okuluva}}, \bibinfo {author}
  {\bibfnamefont {S.}~\bibnamefont {Verkhovskii}}, \bibinfo {author}
  {\bibfnamefont {A.}~\bibnamefont {Yakubovskii}}, \bibinfo {author}
  {\bibfnamefont {L.}~\bibnamefont {Shustov}},\ and\ \bibinfo {author}
  {\bibfnamefont {A.}~\bibnamefont {Trokiner}},\ }\bibfield  {title} {\bibinfo
  {title} {{The $^{63}$Cu and $^{17}$O NMR studies of spin susceptibility in
  differently doped Tl$_2$Ba$_2$CaCu$_2$O$_{8-\delta}$ compounds}},\ }\href
  {https://doi.org/10.1016/s0921-4534(99)00528-6} {\bibfield  {journal}
  {\bibinfo  {journal} {Phys. C Supercond.}\ }\textbf {\bibinfo {volume}
  {328}},\ \bibinfo {pages} {163} (\bibinfo {year} {1999})}\BibitemShut
  {NoStop}%
\bibitem [{\citenamefont {Itoh}\ \emph {et~al.}(2017)\citenamefont {Itoh},
  \citenamefont {Machi},\ and\ \citenamefont {Yamamoto}}]{Itoh2017}%
  \BibitemOpen
  \bibfield  {author} {\bibinfo {author} {\bibfnamefont {Y.}~\bibnamefont
  {Itoh}}, \bibinfo {author} {\bibfnamefont {T.}~\bibnamefont {Machi}},\ and\
  \bibinfo {author} {\bibfnamefont {A.}~\bibnamefont {Yamamoto}},\ }\bibfield
  {title} {\bibinfo {title} {{Ultraslow fluctuations in the pseudogap states
  of}},\ }\href@noop {} {\bibfield  {journal} {\bibinfo  {journal} {Phys. Rev.
  B}\ }\textbf {\bibinfo {volume} {95}},\ \bibinfo {pages} {094501} (\bibinfo
  {year} {2017})}\BibitemShut {NoStop}%
\bibitem [{\citenamefont {Zheng}\ \emph {et~al.}(1996)\citenamefont {Zheng},
  \citenamefont {Kitaoka}, \citenamefont {Asayama}, \citenamefont {Hamada},
  \citenamefont {Yamauchi},\ and\ \citenamefont {Tanaka}}]{Zheng1996}%
  \BibitemOpen
  \bibfield  {author} {\bibinfo {author} {\bibfnamefont {G.-q.}\ \bibnamefont
  {Zheng}}, \bibinfo {author} {\bibfnamefont {Y.}~\bibnamefont {Kitaoka}},
  \bibinfo {author} {\bibfnamefont {K.}~\bibnamefont {Asayama}}, \bibinfo
  {author} {\bibfnamefont {K.}~\bibnamefont {Hamada}}, \bibinfo {author}
  {\bibfnamefont {H.}~\bibnamefont {Yamauchi}},\ and\ \bibinfo {author}
  {\bibfnamefont {S.}~\bibnamefont {Tanaka}},\ }\bibfield  {title} {\bibinfo
  {title} {{{NMR} study of local hole distribution, spin fluctuation and
  superconductivity in
  {Tl}$_{\mathrm{2}}${Ba}$_{\mathrm{2}}${Ca}$_{\mathrm{2}}${Cu}$_{\mathrm{3}}${O}$_{\mathrm{10}}$}},\
  }\href {https://doi.org/10.1016/0921-4534(96)00092-5} {\bibfield  {journal}
  {\bibinfo  {journal} {Physica C: Superconductivity}\ }\textbf {\bibinfo
  {volume} {260}},\ \bibinfo {pages} {197} (\bibinfo {year}
  {1996})}\BibitemShut {NoStop}%
\bibitem [{\citenamefont {Magishi}\ \emph {et~al.}(1995)\citenamefont
  {Magishi}, \citenamefont {Kitaoka}, \citenamefont {Zheng}, \citenamefont
  {Asayama}, \citenamefont {Tokiwa}, \citenamefont {Iyo},\ and\ \citenamefont
  {Ihara}}]{Magishi1995}%
  \BibitemOpen
  \bibfield  {author} {\bibinfo {author} {\bibfnamefont {K.}~\bibnamefont
  {Magishi}}, \bibinfo {author} {\bibfnamefont {Y.}~\bibnamefont {Kitaoka}},
  \bibinfo {author} {\bibfnamefont {G.-q.}\ \bibnamefont {Zheng}}, \bibinfo
  {author} {\bibfnamefont {K.}~\bibnamefont {Asayama}}, \bibinfo {author}
  {\bibfnamefont {K.}~\bibnamefont {Tokiwa}}, \bibinfo {author} {\bibfnamefont
  {A.}~\bibnamefont {Iyo}},\ and\ \bibinfo {author} {\bibfnamefont
  {H.}~\bibnamefont {Ihara}},\ }\bibfield  {title} {\bibinfo {title} {{Spin
  Correlation in High-$T_{\rm c}$ Cuprate
  HgBa$_{2}$Ca$_{2}$Cu$_{3}$O$_{8+\delta}$ with $T_{\rm c}=133\,$K\\ An Origin
  of $T_{\rm c}$-Enhancement Evidenced by $^{63}$Cu-NMR Study}},\ }\href
  {https://doi.org/10.1143/JPSJ.64.4561} {\bibfield  {journal} {\bibinfo
  {journal} {J. Phys. Soc. Japan}\ }\textbf {\bibinfo {volume} {64}},\ \bibinfo
  {pages} {4561} (\bibinfo {year} {1995})}\BibitemShut {NoStop}%
\end{thebibliography}%
\printindex

\newpage
\appendix

\section*{Appendix: Literature Data}
Tab.~\ref{tab:table1} contains the references for all literature data used in this analysis. The values listed in the doping column for single and bilayer materials are those one finds if estimating from the parabolic dependence of \tc on doping alone \cite{Presland1991}, while those listed for multilayer materials are taken as given from the reference where available. UD, OP, and OD in the material labels stand for underdoped, optimally doped, and overdoped, respectively and are followed by the quoted value of \tc. Materials with distinct inner and outer CuO$_2$ planes are labeled additionally with -IP and -OP to signify data pertaining to the inner and outer planes respectively. $T_{\mathrm{metal}}$, which is shown in Fig.~\ref{fig:fig1} is determined from the highest temperature for which the scaled 1/$^{63}T_{1\perp}\geq23.5$/Ks. This value is a lower bound for Hg1201UD72K, since the high-temperature bend in 1/$^{63}T_{1\perp}$ is not visible within the temperature range for which data were available. Planar O quadrupolar splitting can be found in Ref. \cite{Rybicki2016}.\\

\begin{table}[h]
\tiny
\centering
\caption{List of materials, extracted data, and data references used in the analysis. }
\begin{tabular}{|c|c|c|c|c|c|c|c|}
\hline
Material label & Formula & Doping&  Reference for  & Scaling factor for & Relaxation anisotropy & ${^{63}T}_{\mathrm{metal}}$& Reference for\\
&&x&$1/{^{63}T}_1$&1/${^{63}T}_{1\perp}T$&$^{63}R$ ($\pm0.035$)& (K) &$1/{^{17}T}_1$\\
\hline
Hg1201UD72K & HgBa$_2$CuO$_{4+\delta}$ & 0.1 & \cite{Gippius1999}  & 0.81& 3.1 &$\geq 200$& \cite{Bobroff1997} \\
\hline
Tl2201OP85K & Tl$_2$Ba$_2$CuO$_{6+y}$ & 0.16 & \cite{Kambe1993} &-& 1.6 \footnote{relaxation anisotropy $^{63}R$ quoted from \cite{Kambe1993}.} & -&\cite{Kambe1993} \\
\hline
Tl2201OD72K & Tl$_2$Ba$_2$CuO$_{6+y}$ & 0.2  & \cite{Fujiwara1991} &1.18 & 1.4 &116& - \\
\hline
Tl2201OD40K & Tl$_2$Ba$_2$CuO$_{6+y}$ & 0.24  & \cite{Fujiwara1991}& 1.23 &1.3 &69 & - \\
\hline
Tl2201OD0K & Tl$_2$Ba$_2$CuO$_{6+y}$ & 0.27  & \cite{Fujiwara1991} & 1.1 & 1 &14& \cite{Kambe1991} \\
\hline
Tl1212OD70K & TlSr$_2$CaCu$_2$O$_{7-\delta}$ & 0.21  & \cite{Magishi1996} & 0.94& 1.5&117 & -\\
\hline
Tl1212OD52K & TlSr$_2$CaCu$_2$O$_{7-\delta}$ & 0.23  & \cite{Magishi1996}&0.99& 1.5  &104& -\\
\hline
Tl1212OD10K & TlSr$_2$CaCu$_2$O$_{7-\delta}$ & 0.265  & \cite{Magishi1996}& 1.15& 1.25 &58& -\\
\hline
YBCOUN62 & YBa$_2$Cu$_3$O$_{6.63}$ & 0.11& \cite{Takigawa1991}&1.12 & 3.6 &225& \cite{Takigawa1991}\\
\hline
YBCO6.92 & YBa$_2$Cu$_3$O$_{6.92}$ & 0.14 &  \cite{Auler1999}& 1.14 & 3.2&158 &-\\
\hline
YBCO6.96 & YBa$_2$Cu$_3$O$_{6.96}$ & 0.16 &  -&- &- &-&\cite{Yoshinari1992}\\
\hline
YBCO7 & YBa$_2$Cu$_3$O$_7$ & 0.17 & \cite{Walstedt1989} &1.22& 2.7 &117& \cite{Kitaoka1989}\\
\hline
Y1248 & YBa$_2$Cu$_4$O$_8$ & 0.12 & \cite{Zimmermann1991} &1.21& 3.3 &214& \cite{Bankay1994}\\
\hline
Tl2212UN102K & Tl$_2$Ba$_2$CaCu$_2$O$_{8-\delta}$ & 0.127 &  \cite{Gerashchenko1999} \footnote { $1/{^{63}T}_{1\perp}$ not measured directly but deduced from spin echo decay.\label{footnoteb}} &0.64& 2.8 \ref{footnoteb}&160 & \cite{Gerashchenko1999}\\
\hline
Tl2212OP112K & Tl$_2$Ba$_2$CaCu$_2$O$_{8-\delta}$ & 0.16 &  \cite{Gerashchenko1999} \ref{footnoteb} &0.86& 2.5 \ref{footnoteb} &180& \cite{Gerashchenko1999}\\
\hline
Tl2212OD104K & Tl$_2$Ba$_2$CaCu$_2$O$_{8-\delta}$ & 0.189 & \cite{Gerashchenko1999} \ref{footnoteb} &0.71 & 2.4 \ref{footnoteb} &180& \cite{Gerashchenko1999}\\
\hline
Hg1212OP127K & HgBa$_2$CaCu$_2$O$_{6+\delta}$ & 0.16 & \cite{Itoh2017}&1.15 & 1.95 &227& -\\
\hline
Tl2223OD115K-IP & Tl$_2$Ba$_2$Ca$_2$Cu$_3$O$_{10-\delta}$ & 0.124 & \cite{Zheng1996} &1.07& 1.85&180&  \cite{Zheng1996}\\
\hline
Tl2223OD115K-OP & Tl$_2$Ba$_2$Ca$_2$Cu$_3$O$_{10-\delta}$ & 0.196 & \cite{Zheng1996}& 1.04 &  1.8 &180& \cite{Zheng1996}\\
\hline
Hg1223OP133K-IP & HgBa$_2$Ca$_2$Cu$_3$O$_{8-\delta}$ & 0.14 &  \cite{Magishi1995}&1.01 & 2 &199& \cite{Magishi1995}\\
\hline
Hg1223OP133K-OP & HgBa$_2$Ca$_2$Cu$_3$O$_{8-\delta}$ & 0.19 &  \cite{Magishi1995} &1.13 & 1.9 &189& \cite{Magishi1995}\\
\hline

\end{tabular}

\label{tab:table1}
\end{table}

\end{document}